\documentclass[12pt,preprint]{aastex}
\newcommand{\reff}{\mbox{$R_{\rm eff}$}}
\newcommand{\feh}{\mbox{[Fe/H]}}

\newcommand{\vi}{\mbox{$V\!-\!I$}}

\begin{document}
\title{On the size difference between red and blue globular clusters
 \footnote{Based on observations with the NASA/ESA Hubble Space 
 Telescope, obtained at the Space Telescope Science Institute, which is 
 operated by the Association of Universities for Research in Astronomy, 
 Inc.\ under NASA contract No.\ NAS5-26555.}
}
\author{S{\o}ren S.\ Larsen\altaffilmark{1} and Jean P.\ Brodie}
\affil{UC Observatories / Lick Observatory, University of California,
       Santa Cruz, CA 95064, USA}
\email{slarsen@eso.org and brodie@ucolick.org}
\altaffiltext{1}{present address: European Southern Observatory, 
  Karl-Schwarzschild Str.\ 2, 85748 Garching b.\ M{\"u}nchen, Germany}

\begin{abstract}
  Several recent studies have reported a mean size difference of about 20\% 
between the metal-rich and metal-poor subpopulations of globular clusters (GCs) 
in a variety of galaxies. In this paper we investigate the possibility that 
the size difference might be a projection effect, resulting from a correlation 
between cluster size and galactocentric distance, combined with different 
radial distributions of the GC subpopulations. We find that projection effects 
may indeed account for a size difference similar to the observed one,
provided that there is a steep relation between GC size and galactocentric
distance in the central parts of the GC system and that the density of GCs 
flattens off near the center in a manner similar to a King profile. For 
more centrally peaked 
distributions, such as a de Vaucouleurs $R^{1/4}$ law, or for shallower
size-radius relations, projection effects are unable to produce the observed
differences in the size distributions. 
\end{abstract}


\section{Introduction}

  Since the work of \citet{kin59} and \citet{zin85}, it has been known
that globular clusters in the Milky Way can be divided into (at least) two
sub-populations with distinct kinematical and chemical properties.  Undoubtedly 
one of the most important developments in research on globular clusters 
(GCs) within the last decade is the discovery that similar substructure
is seen in the color (and hence, presumably, metallicity) distributions of
GC systems in many \emph{early-type} galaxies \citep{za93,gk99,kw01,lar01}. 
There is increasing evidence that GC sub-populations in early-type galaxies 
share many properties with those in spirals \citep{fbl01}, and that the
GCs in different galaxy types may be closely related.  Characterizing and
understanding the properties (spatial and metallicity distributions, 
kinematics etc.) of GC sub-populations is currently a subject of much 
investigation and holds the promise of revealing important information about 
events in the evolution of their host galaxies.

  One piece of the puzzle is to establish just how similar are GCs in
different galaxies.  The Hubble Space Telescope can
resolve the spatial profiles of globular clusters well beyond 
the Local Group, although careful modeling of the undersampled point spread 
function (PSF) of the WFPC2 camera is necessary in order to derive reliable 
size information for typical extragalactic GCs.  \citet{kw98} were among the 
first to measure the sizes of 
extragalactic GCs and found that GCs in the lenticular galaxy NGC~3115 had 
half-light radii of about 2 pc, similar to or perhaps slightly smaller than 
those in the Milky Way.  When measuring sizes for the blue (metal-poor) and 
red (metal-rich) GC subpopulations separately, \citet{kw98} noted a size 
difference of about 20\% with the red GCs being systematically smaller.  A 
similar size difference was found between red and blue GCs in NGC~4486 
\citep{kun99}.  Subsequently, size differences between GC subpopulations have 
been found in many other 
galaxies including NGC~4472 \citep{puz99}, NGC~4594 
\citep[the ``Sombrero''; ][]{lfb01}, 
M31 \citep{bhh02} and other early-type galaxies \citep{lar01}. However, it 
should also be noted that \citet{har02} found \emph{no} size-color relation 
for a sample of 27 GCs in NGC~5128. 

  Although the results on NGC~5128 suggest that the size difference may not be 
\emph{universal}, there seems to be compelling evidence that it is at least a 
wide-spread phenomenon, observed in spirals as well as in early-type galaxies. 
Understanding the origin of this size difference is a high priority, 
since any intrinsic correlation between the sizes and metallicities of star 
clusters might hold clues to their formation mechanisms. However, before one 
attempts to explain the size difference in terms of e.g.\ the properties of 
the proto-cluster clouds, variations in external pressure etc., other more
straight-forward explanations need to be ruled out. 

  One possibility is that the observed size difference is a projection 
effect, resulting from a correlation between GC size ($r$) and
Galactocentric distance ($R$) combined with different radial distributions
of the GC subpopulations (throughout this paper, we will adopt the 
following convention: a small $r$ refers to the size (radius) of an individual 
globular cluster, while a capital $R$ refers to the distance of that cluster 
from the galaxy center. We will use subscripts ($R_{\rm proj}$, $R_{\rm 3D}$) 
to specify projected and 3-D quantities).
\citet{van91} found that the size-$R$ relation 
in the Milky Way can be approximated by a square-root relation, 
$r \propto \sqrt{R}$, and they also noted that similar relations exist in M31, 
NGC~5128 and the LMC. However, because of the limited spatial coverage of 
HST, little is known about size-$R$ trends in other more distant galaxies.  

  Concerning the wide-field spatial distributions of extragalactic GC systems,
most of the available information comes from ground-based studies which 
generally provide the only way to obtain complete coverage at large radii.
In addition to the Milky Way and M31, galaxies where the radial distributions 
of GC sub-populations have been studied include NGC~1380 \citep{kis97}, 
NGC~1399 \citep{dir03}, NGC~4472 \citep{lee98} and NGC~4486 \citep{cot01}. 
In all of these cases, the red GC sub-population is more centrally 
concentrated than the blue one.
Therefore, red GCs observed at a given \emph{projected} radius will
on average be physically closer to the galaxy center than the
blue ones.  Clearly, there is reason to suspect that projection effects might 
account for at least some of the differences in the mean sizes of GC 
samples.  Our goal in this paper is to investigate and quantify this idea 
in more detail.

\section{Observations: sizes of GC subpopulations}

  While size information is now available for many extragalactic GC systems, 
the spatial
coverage is in most cases limited to a single WFPC2 pointing, usually
centered on the galaxy. Therefore, information about variations with
galactocentric distance is limited. Even with several pointings, the
surface density of GCs drops off rapidly at large radii, and in all but
the richest systems it is difficult to obtain reliable statistics based on 
the small number of clusters contained within an off-center WFPC2 field. 
In practice, then, the 
number of suitable datasets that can serve as a basis for our analysis is 
limited.  We have selected three well-studied galaxies with multiple 
HST/WFPC2 pointings and rich GC systems (NGC~4472, NGC~4486 and NGC~4594). 
The WFPC2 data are the same as those used in \citet{lar01} and we refer to 
that paper for details about the data reduction. Briefly, the globular
cluster sizes were measured by convolving a series of analytic models
with the WFPC2 PSF, adjusting the FWHM of the model until the best
match to the observed profile was obtained. For this purpose, we used
the ISHAPE code described in \citet{lar99}. The main systematic errors 
involved in this process are related to che choice of model PSF and are
difficult to quantify, but comparison of size measurements on different 
frames have indicated that this method yields FWHM values consistent to 
about 0.1 WF pixel for individual clusters \citep{lfb01,lar01}. In addition, 
we include data for the Milky Way from \citet{har96} and M31 
(size information kindly provided by P.\ Barmby).  

\begin{figure}
\begin{minipage}{52mm}
\plotone{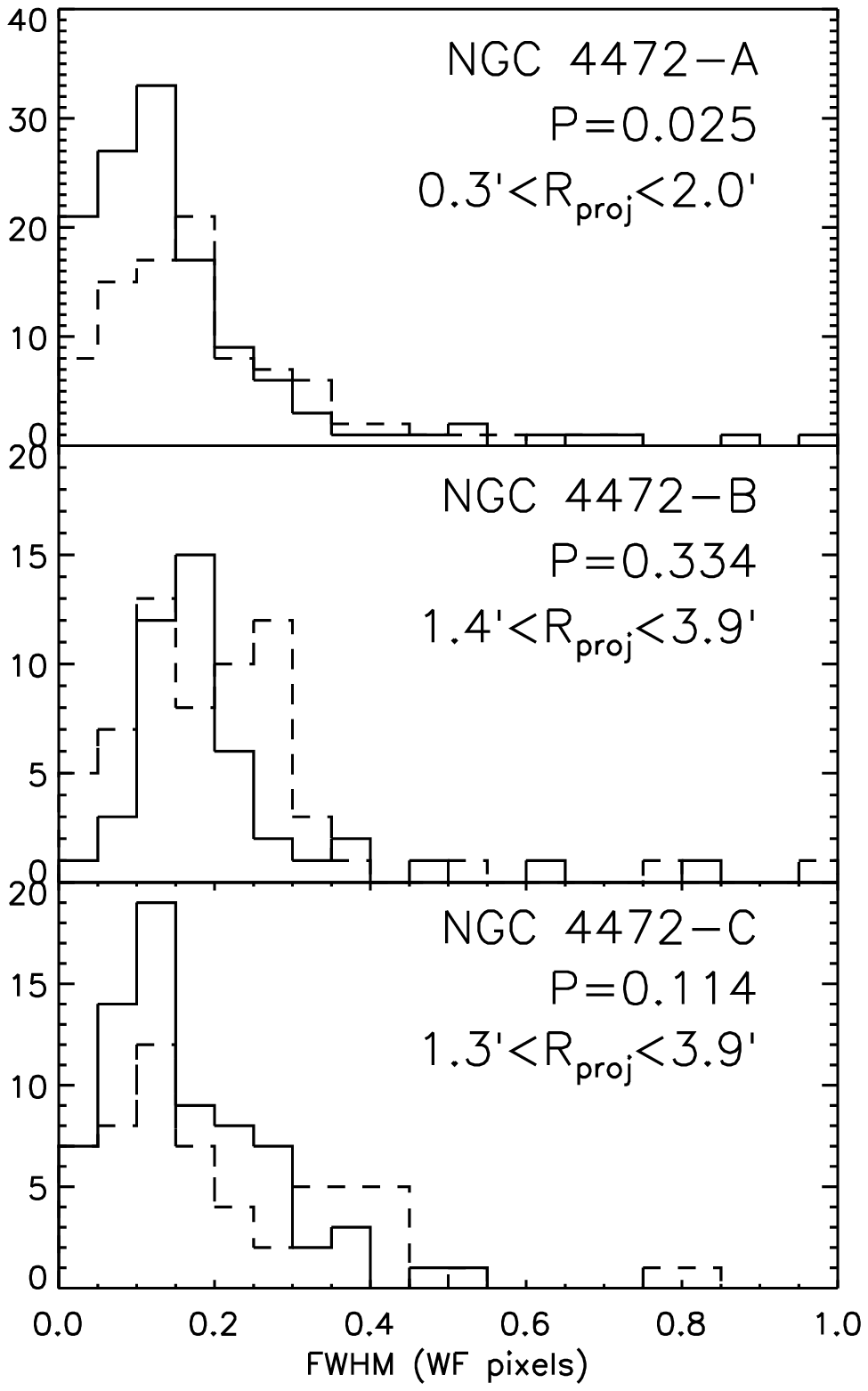}
\end{minipage}
\begin{minipage}{52mm}
\plotone{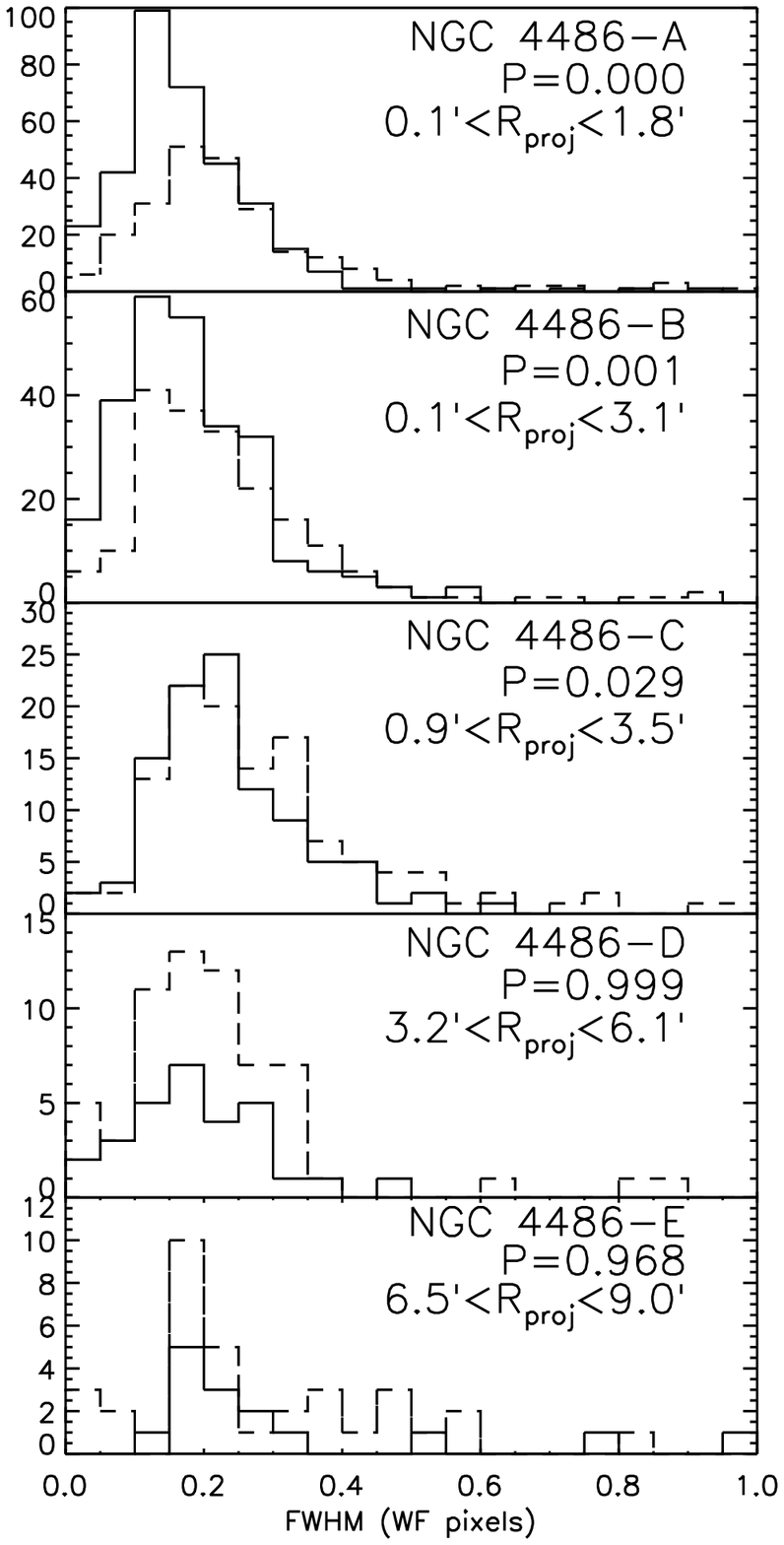}
\end{minipage}
\begin{minipage}{52mm}
\plotone{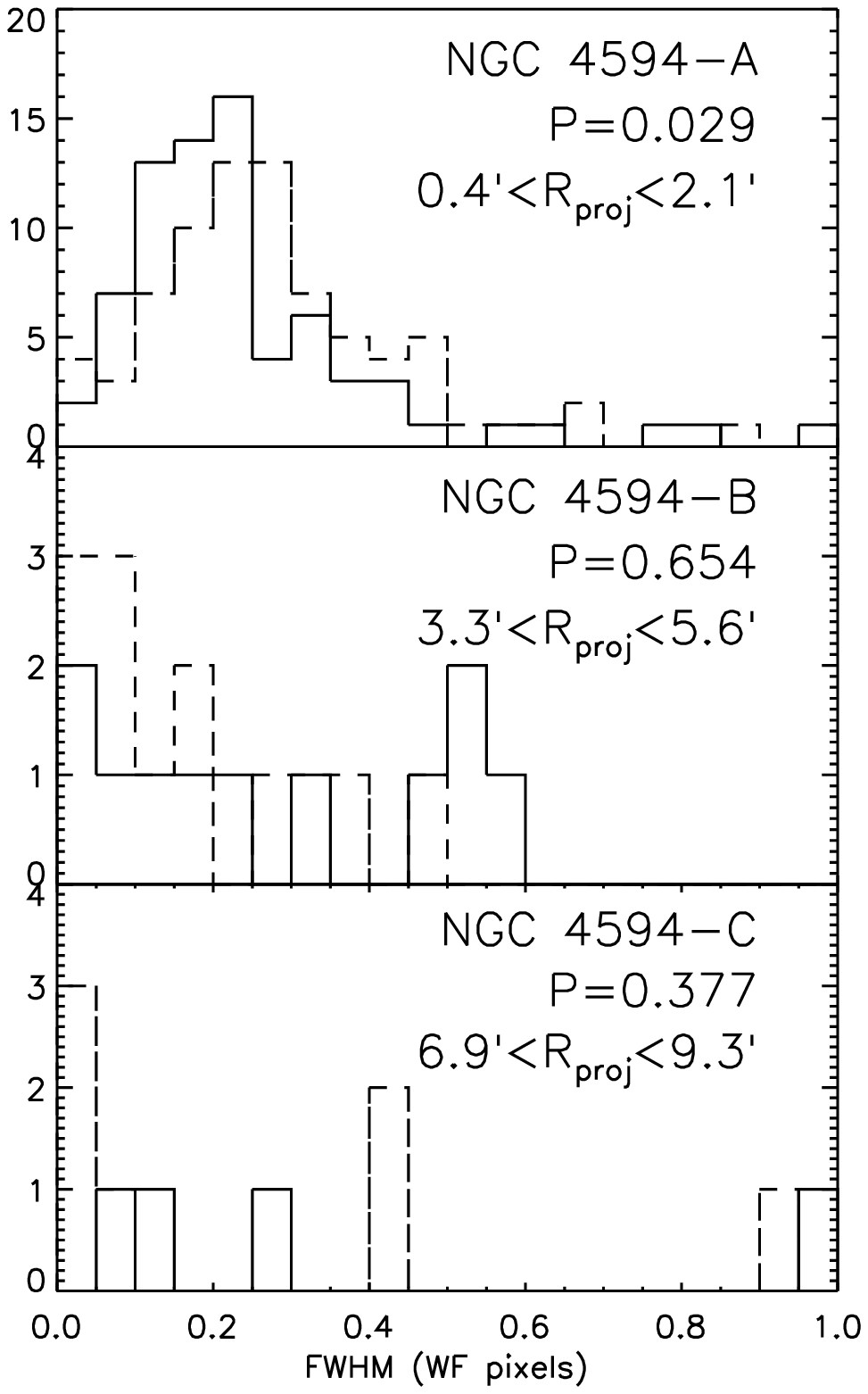}
\end{minipage}
\figcaption[]{\label{fig:szdist}Size distributions for globular clusters
  in NGC~4472, NGC~4486 and NGC~4594. Red ($\vi>1.05$) and blue GCs are
  shown with solid and dashed lines, respectively. The $P$ values are
  the probability from a Kolmogorov-Smirnoff test that the two distributions
  are drawn from the same parent distribution. Each panel contains data
  for one WFPC2 pointing (WF chips only).
}
\end{figure}

  As an illustration of the typical size differences between GC 
sub-populations, Fig.~\ref{fig:szdist} shows the size distributions for GCs 
in each of the WFPC2 pointings for the three early-type galaxies.  Here we 
are not concerned with the absolute 
values of the GC sizes so for the HST data we simply use the FWHM of 
the cluster profiles, measured in pixels and corrected for the WFPC2 PSF.
To avoid any systematic differences due to the different pixel scales of the 
PC and WF chips, we only use sizes measured on the WF chips.  We divide
between ``red'' and ``blue'' clusters at $\vi = 1.05$, corresponding roughly
to a metallicity of $\feh\sim-1$.
  Table~\ref{tab:sztab} lists the mean sizes of red and blue globular 
clusters in each of the WFPC2 fields, along with the probability $P$ (from a 
Kolmogorov-Smirnoff test) that the size distributions are drawn from the 
same parent distribution.  For all of the \emph{central} WFPC2 pointings, 
Table~\ref{tab:sztab} and Fig.~\ref{fig:szdist} confirm a significant
difference between the size 
distributions of blue and red GCs ($P<0.03$), amounting to a
difference of 15--30\% in the mean sizes.  However, at larger radii the 
situation is less clear: While the red GCs are smaller than the blue GCs in
all of the NGC~4486 pointings except NGC~4486-E, the difference is only 
significant in the three innermost ones. In 
NGC~4472, the two outer pointings (B and C) cover roughly the same radial 
range, but a size difference is seen in only one of them and is only 
significant at the $\sim$89\% confidence level.  In NGC~4594, only the 
central pointing has enough clusters 
to obtain useful information about size distributions. In this central
pointing there is a $\sim15$\% difference between the mean sizes of red
and blue GCs and the distributions are different at the 97\% confidence level.

\begin{figure}
\begin{minipage}{12cm}
\plotone{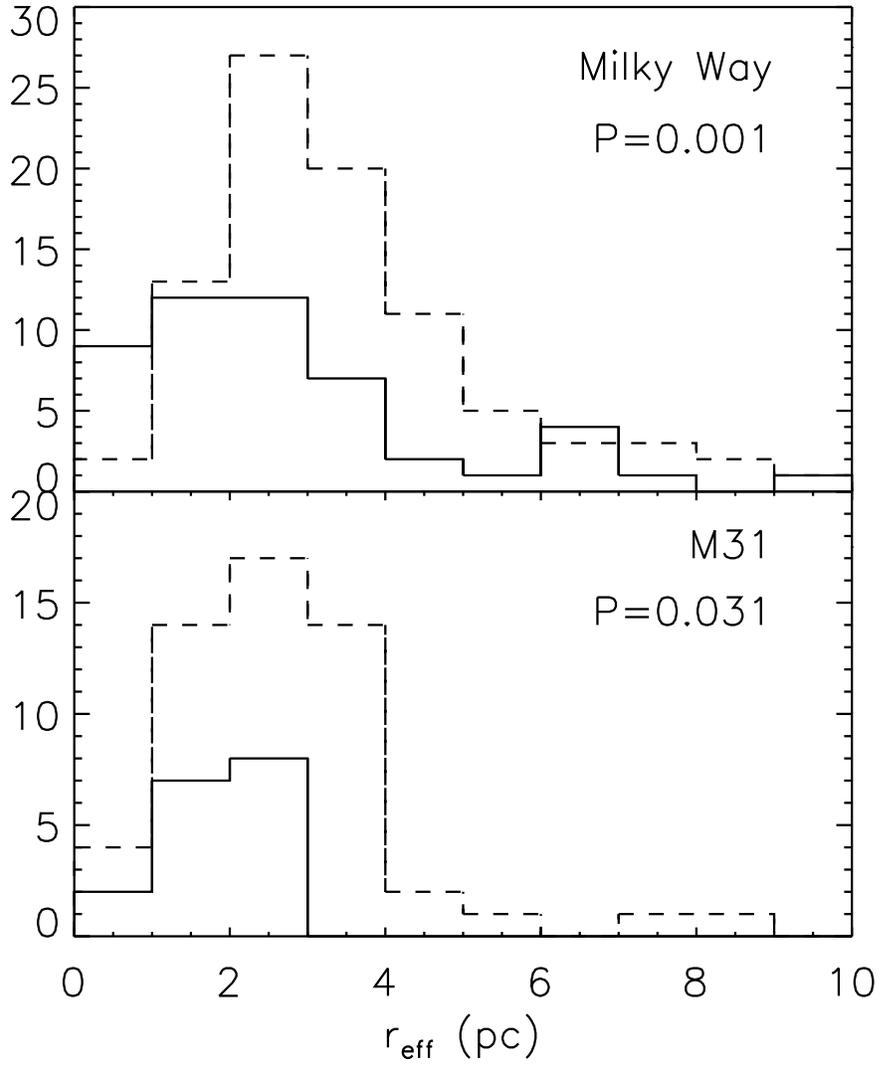}
\end{minipage}
\figcaption[]{\label{fig:szdist_sp}Same as Fig.~\ref{fig:szdist}, but for
  Milky Way and M31 globular clusters. Data for M31 GCs kindly provided
  by P.\ Barmby.
}
\end{figure}

  The size distributions for Milky Way and M31 GCs are shown in 
Fig.~\ref{fig:szdist_sp}.  In the Milky Way we have adopted a division 
at $\feh=-1.0$,
while the classification of clusters in M31 as metal-poor or metal-rich was 
based on a variety of photometric and spectroscopic criteria \citep{barm00}.  
As previously noted \citep{lar01,bhh02}, metal-rich and metal-poor GCs in 
the Milky Way and M31 display a size difference reminiscent of that 
observed in the early-type galaxies, but at least in the Milky Way it is 
likely that the difference is at least partly due to the different radial 
distributions of the two GC subpopulations. In \citet{lar01} we tried to 
reduce this problem by comparing clusters in smaller radial bins, but the 
modest number of GCs in the Milky Way limits the extent to which such 
sub-division is feasible. In Sec.~\ref{ssec:rev_mw} below we return to
this issue.

\section{Modeling projection effects}

\subsection{GC size versus luminosity}

  Before we proceed, we briefly discuss one other effect which could 
potentially affect the comparison of mean sizes for GC subpopulations. Because 
the mass-to-light ratio of a single-aged stellar population (such as a star
cluster) is a function of metallicity, the mass limit of a 
magnitude-limited sample of GCs is, in principle, also metallicity-dependent.
Thus, a size-mass relation could introduce spurious size
differences between magnitude-limited samples of different metallicities.
Furthermore, age differences between GC sub-populations could also lead to 
different mass-to-light ratios. However, in the Milky Way there is a 
striking \emph{absence} of any size-luminosity relation for GCs \citep{van91},
and the same seems to be the case in early-type galaxies \citep{kw01} and even 
for young clusters in mergers \citep{zepf99}. This suggests that the magnitude 
(or mass) limit of the sample is largely irrelevant when comparing 
cluster sizes. 

\begin{figure}
\begin{minipage}{12cm}
\plotone{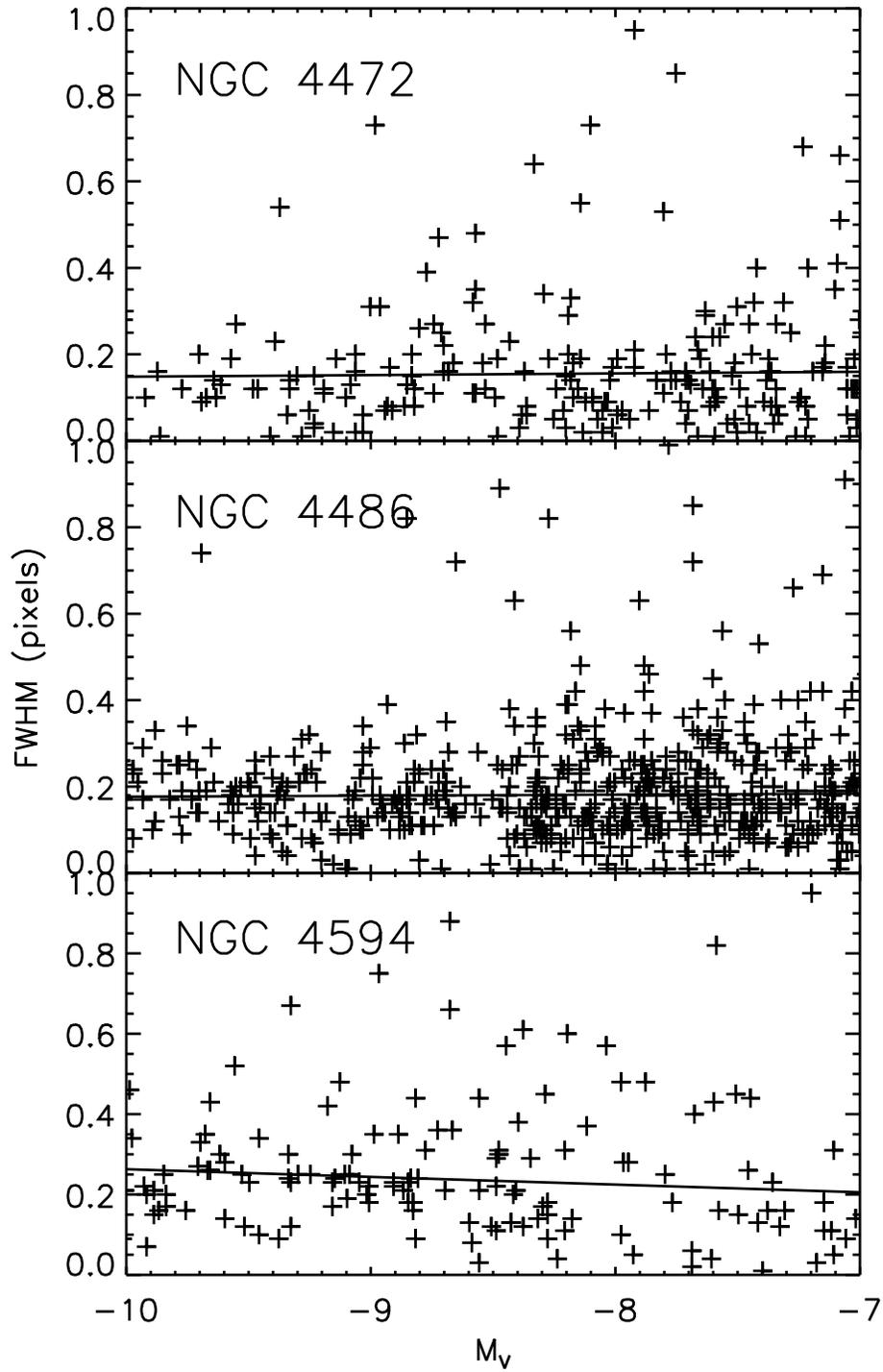}
\end{minipage}
\figcaption[]{\label{fig:sz_mag}Size vs.\ magnitude for GCs in NGC~4472,
  NGC~4486 and NGC~4594. The lines are linear fits to the data,
  with slopes of $0.004\pm0.009$ pixels mag$^{-1}$ (NGC~4472), 
  $0.002\pm0.005$ pixels mag$^{-1}$ (NGC~4486)
  and $-0.019\pm0.014$ pixels mag$^{-1}$ (NGC~4594).
}
\end{figure}

  In Fig.~\ref{fig:sz_mag} we plot the size measurements for globular clusters 
in NGC~4472, NGC~4486 and NGC~4594 as a function of their magnitude.  The 
lines are least squares linear fits to the data, and they all have very 
small slopes that differ from 0 by less than $1\sigma$. Again, this suggests 
that the size difference cannot be explained as a result of different M/L 
ratios of the GC sub-populations.
  We note in passing that this lack of a size-luminosity relation is a very 
interesting and puzzling result in its own right, with the obvious implication 
that the mean \emph{densities} of stellar clusters are directly proportional 
to their masses.

\subsection{Revisiting the Milky Way}
\label{ssec:rev_mw}

  In order to calculate the projected size distributions of GC sub-populations, 
two basic ingredients are needed: 1) a relation between GC size and 
galactocentric distance, assumed to be \emph{the same} for both (or all) GC 
subpopulations, and 2) density profiles, assumed to be 
\emph{different} for each population. The size-$R$ relation is not easily 
constrained for extragalactic GC systems, partly because the GC sizes are 
close to the resolution limit of WFPC2 and small differences in telescope 
tracking, focusing etc.\ could introduce spurious effects when comparing 
measurements from different pointings. A more fundamental problem is that only 
the projected size-$R$ relations can be directly observed, while we need the 
run of GC size versus physical (3-D) galactocentric distance. For the
following discussion we will therefore start out by \emph{assuming} a 
relation of the form
\begin{equation}
  r \, = \, a \sqrt{R_{\rm 3D}} \, + \, b,
  \label{eq:r_R}
\end{equation}
 i.e.\ similar to that observed in the Milky Way, but note that we have 
allowed for addition of a constant term to the simple square-root law 
suggested by \citet{van91}.  
We emphasize 
that we do not intend to imply any physical significance of this particular 
relation but simply use it as a convenient fitting function.  Note, also, 
that any intrinsic scatter in the relation (which might depend on $R_{\rm 3D}$
in non-trivial ways) is ignored.

\begin{figure}
\begin{minipage}{11cm}
\plotone{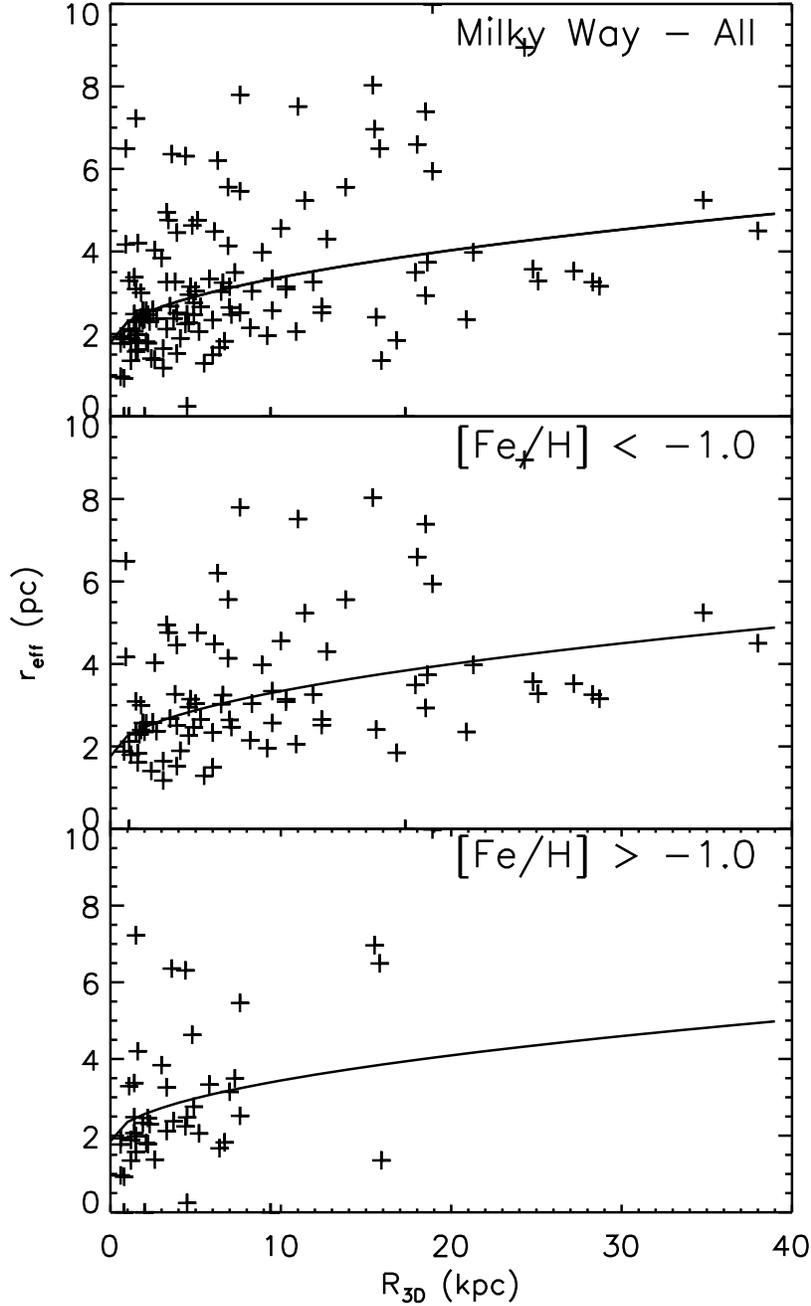}
\end{minipage}
\figcaption[]{\label{fig:r_sz_mw}Globular cluster size vs.\ distance from
 Milky Way center for Galactic globular clusters.  The solid line in the 
 upper panel is a fit of the 
 form $r_{\rm eff} = a \, \sqrt{R_{\rm 3D} / 3 \, {\rm kpc}} \, + \, b$ with
 $a=0.87\pm0.16$ pc and $b = 1.79\pm0.26$ pc. In the two lower panels the 
 slope $a$ was kept fixed. Solving for $b$ alone yields
 $b = 1.76\pm0.13$ pc for the metal-poor clusters and $b=1.86\pm0.23$ pc for
 the metal-rich ones.
}
\end{figure}

  Fig.~\ref{fig:r_sz_mw} shows $r_{\rm eff}$ versus 
$R_{\rm 3D}$ for Milky Way GCs.  
The upper panel shows a least-squares fit of 
Eq.~(\ref{eq:r_R}) to all clusters with size information. The best fit is 
\begin{equation}
  r_{\rm eff} = (0.87\pm0.16 \, {\rm pc}) \sqrt{R_{\rm 3D} / 3 {\rm kpc}} \, 
   + \, (1.79\pm0.26 \, {\rm pc}),
  \label{eq:r_R_mw}
\end{equation}
where $r_{\rm eff}$ is the effective (half-light) radius and we have normalized
the relation to a scale radius of 3 kpc (roughly the median galactocentric
distance of the metal-rich GCs). Of course, choosing a different scale
radius is equivalent to changing the coefficient $a$.  The two lower panels 
show the metal-poor
(``blue'') and metal-rich GCs separately. Especially for the metal-rich GCs, 
the slope of the size-$R$ relation is very poorly constrained, but to 
check whether any significant size differences exist between the two GC 
populations we 
adopt the same slope as for the combined sample and just fit the constant 
term. For the metal-poor and metal-rich GCs this term is $b = 1.76\pm0.13$ pc 
and $b = 1.86\pm0.23$ pc, respectively.  Formally, the constant term is now
slightly larger for the \emph{metal-rich} clusters, contrary to the trend 
seen in the projected size distributions, but the difference is not 
statistically significant. In practice, then, there is no evidence for a
significant size 
difference between the two Milky Way GC populations when the radial
dependence is averaged out. We therefore conclude that the size
difference in Fig.~\ref{fig:szdist_sp}, at least for the Milky Way, 
can be fully explained as a result of the different radial distributions of 
the two GC subpopulations.

\subsection{Extragalactic GC systems}

\begin{figure}
\begin{minipage}{12cm}
\plotone{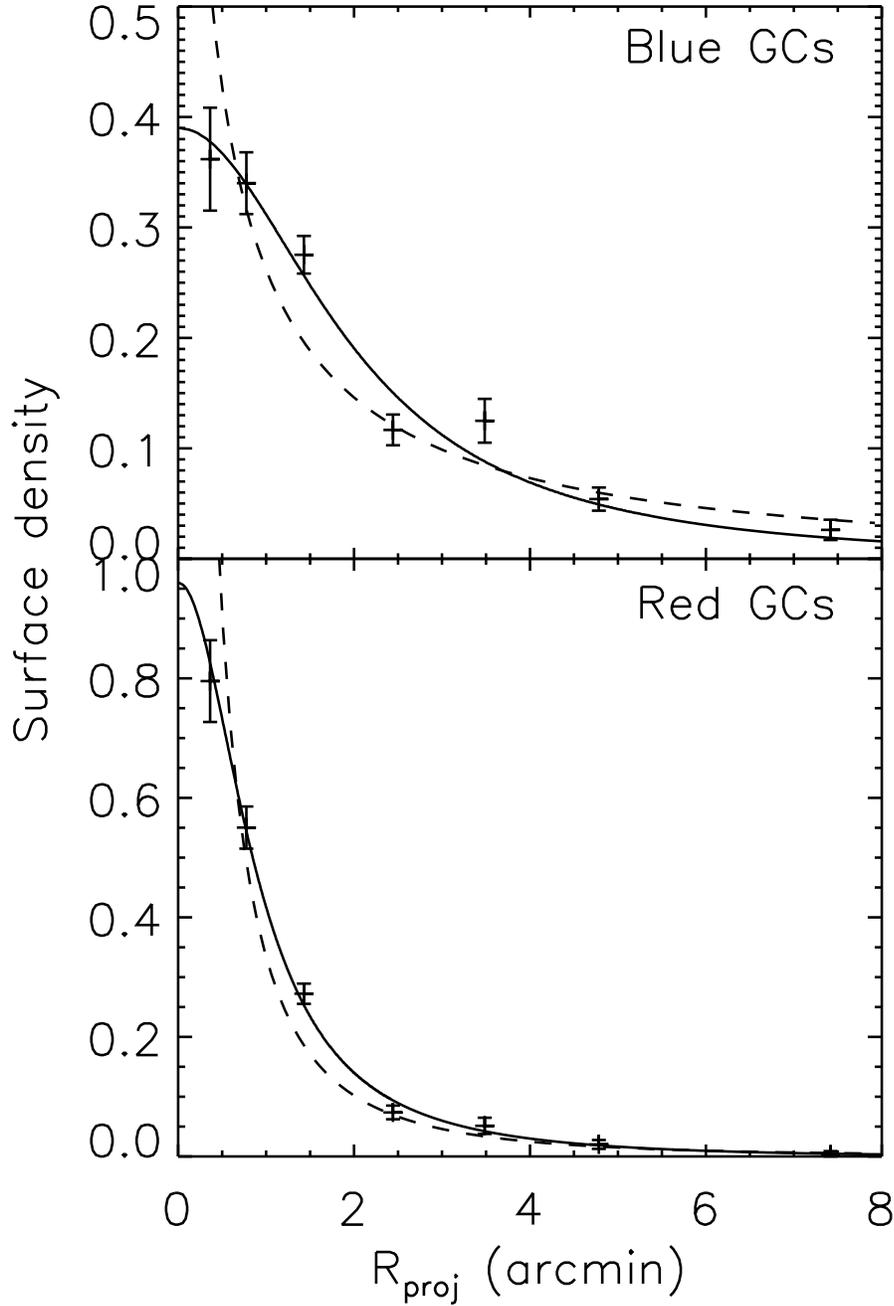}
\end{minipage}
\figcaption[]{\label{fig:rp4486}Surface densities of blue (top) and
  red (bottom) GCs in NGC~4486, from HST/WFPC2 pointings. The solid and
  dashed lines represent least-squares King and de Vaucouleurs profile fits 
  to the data. Note that the best-fitting de Vaucouleurs profile to
  the blue GCs has a much larger effective radius than the corresponding
  best-fitting King profile.
}
\end{figure}

  While ground-based data offer superior coverage at large radii compared to
most HST datasets, completeness issues are more severe near the center. 
However, in order to model the GC size distributions it is important to
know the behavior of the spatial distributions of GCs near the center,
as well as at larger distances. Of the HST datasets used here, NGC~4486
has the best radial coverage, extending out to 
9$\arcmin$.  In Fig.~\ref{fig:rp4486} we show the GC surface densities as
a function of projected distance from the galaxy center (in arcmin)
for the two GC subpopulations.  It is evident that the red GCs are much 
more centrally concentrated.  We have adopted a magnitude limit of
$V=24$, which is 1.5 mag brighter than the 50\% completeness limit even
in the central pointing \citep{lar01}, so we do not expect completeness
effects to be significant except perhaps in the central 
20$\arcsec$--30$\arcsec$.
Also superimposed on the plots are de Vaucouleurs 
$R^{1/4}$ laws and 
\citet{king66} profiles.  For the King profiles we have arbitrarily fixed the 
concentration parameter at $R_{\rm tidal}/R_{\rm core} = 20$.
The two best-fitting de Vaucouleurs profiles have effective radii
of $2\arcmin$ (red clusters) and $24\arcmin$ (blue clusters), although 
these numbers are rather sensitive to the background correction.
The two King profiles have effective radii of $2\farcm05$ and $4\farcm56$,
a much smaller difference than for the de Vaucouleurs profiles, but note
that our radial coverage only extends out to about 1/3 of the half-light
radius of the de Vaucouleurs fit to the blue GC distribution. 

In the
outer, poorly constrained regions, the King profiles decline more rapidly 
than the $R^{1/4}$ law, resulting in the very different half-light radii for 
the two types of profiles.  In fact, a fundamental difference between
the King and de Vaucouleurs profiles is that the former has a
well-defined tidal radius, while the latter does not.
However, our primary objective here is not to obtain a correct fit
at very large galactocentric distances (which contribute with only a few 
clusters in projection),
but merely to find appropriate ``fitting functions'' that allow us
to model the spatial distributions of GCs and deproject the observed surface
densities to obtain estimates of 3-D space densities.  Fig.~\ref{fig:rp4486} 
suggests that the King profiles provide a fit that is at least as good as an
$R^{1/4}$ law in the inner regions, and probably better. 

  One remaining issue is to convert the observed surface density profiles
to space densities, $\rho(R_{\rm 3D})$.  For the de Vaucouleurs profiles we 
use the series 
expansion tabulated by \citet{ben93}. For the King profiles, the space
density is automatically obtained during the process of computing the King 
model. 

\subsection{Projected mean properties of GC sub-populations}

  Near the center, the King profiles flatten out whereas the de
Vaucouleurs profiles diverge.
This difference in behavior has important implications for the projected 
properties of GC
systems distributed according to either of the two profiles. To illustrate
this point, consider the mean 3-D distance $\langle R_{\rm 3D} \rangle$ of 
GCs observed at a given projected distance $R_{\rm proj}$. These are
related as
\begin{equation}
  \langle R_{3D}\rangle_{R_{\rm proj}} \, = \, 
    \frac{
      \int_{R_{\rm proj}}^{R_{\rm max}} R_{\rm 3D} \, n(R_{\rm 3D}) dR_{\rm 3D}
    }{
      \int_{R_{\rm proj}}^{R_{\rm max}} 
        n(R_{\rm 3D}) dR_{\rm 3D}
    }
  \label{eq:ravg}
\end{equation}
where
\begin{equation} 
  n(R_{3D}) \equiv \frac{R_{3D} \rho(R_{3D})}{\sqrt{R_{3D}^2 - R_{\rm proj}^2}}
  \label{eq:ns}
\end{equation}
  The upper integration limit, $R_{\rm max}$, should in principle be set to 
the outer
radius of the GC system. This is a poorly constrained, but relatively 
uncritical number as the density in the outer parts is very low.
We will generally adopt $R_{\rm max} = 20 \, \reff_R$,
where $\reff_R$ is the effective radius of the red GC subpopulation. This
is well beyond the limit where reliable data exists for most GC systems.

\begin{figure}
\begin{minipage}{11cm}
\plotone{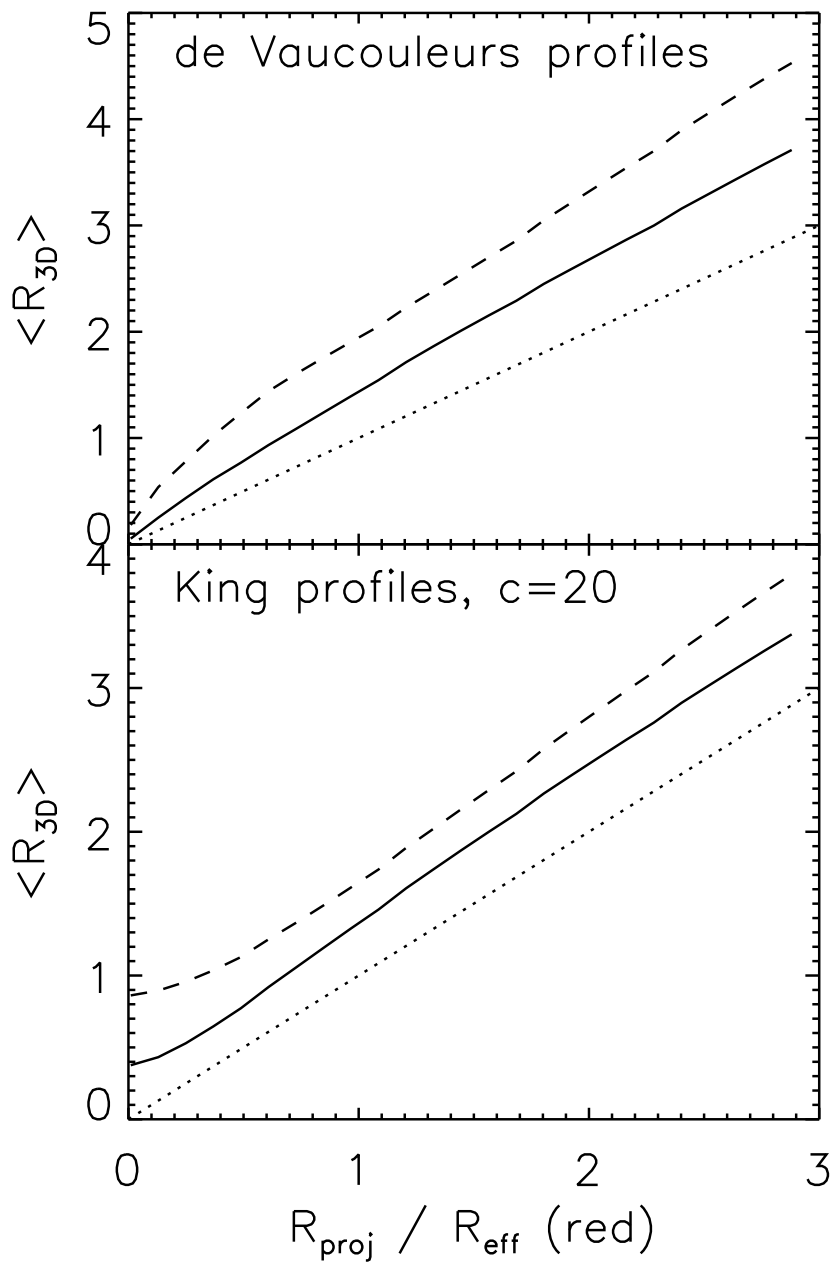}
\end{minipage}
\figcaption[]{\label{fig:ipar_r}Mean 3-D radial distance versus projected
  distance for de Vaucouleurs (top) and King (bottom) profiles.  Dotted 
  lines represent a 1:1 relation, solid lines are for a distribution with 
  $\reff$ = 1 (``red GCs'') and the dashed lines are for $\reff = 10$ 
  (de Vaucouleurs profiles) or $\reff = 2.3$ (King profiles).
  cf.\ Fig.~\ref{fig:rp4486}. 
}
\end{figure}

  Fig.~\ref{fig:ipar_r} shows $\langle R_{3D}\rangle$ as a function of 
$R_{\rm proj}/\reff_R$ for de Vaucouleurs and King profiles corresponding
to the fits in Fig.~\ref{fig:rp4486}.  The straight dotted line in each 
figure is simply a 1:1 relation, plotted for reference.
Fig.~\ref{fig:ipar_r} shows that, at any $R_{\rm proj}$, the ``blue'' GCs 
are indeed located at significantly larger mean distances from the galaxy
center than the red ones.  For the de Vaucouleurs profile, most of the 
clusters observed near 
the center are also located at small physical distances from the center.  
For the King profiles, which flatten out near the center, the mean 3D 
distance from the center at $R_{\rm proj}=0$ is substantially larger 
than for the de Vaucouleurs profiles.

  In a similar fashion to Eq.~(\ref{eq:ravg}), the mean cluster size 
$\langle r \rangle$ at $R_{\rm proj}$ is
\begin{equation}
  \langle r \rangle_{R_{\rm proj}} \, = \, 
    \frac{
    \int_{R_{\rm proj}}^{R_{\rm max}} 
      r(R_{\rm 3D}) n(R_{\rm 3D}) dR_{\rm 3D}
    }{
      \int_{R_{\rm proj}}^{R_{\rm max}} 
        n(R_{\rm 3D}) dR_{\rm 3D}
    }
  \label{eq:szavg}
\end{equation}

\subsection{Application to the M31 GC system}

\begin{figure}
\begin{minipage}{12cm}
\plotone{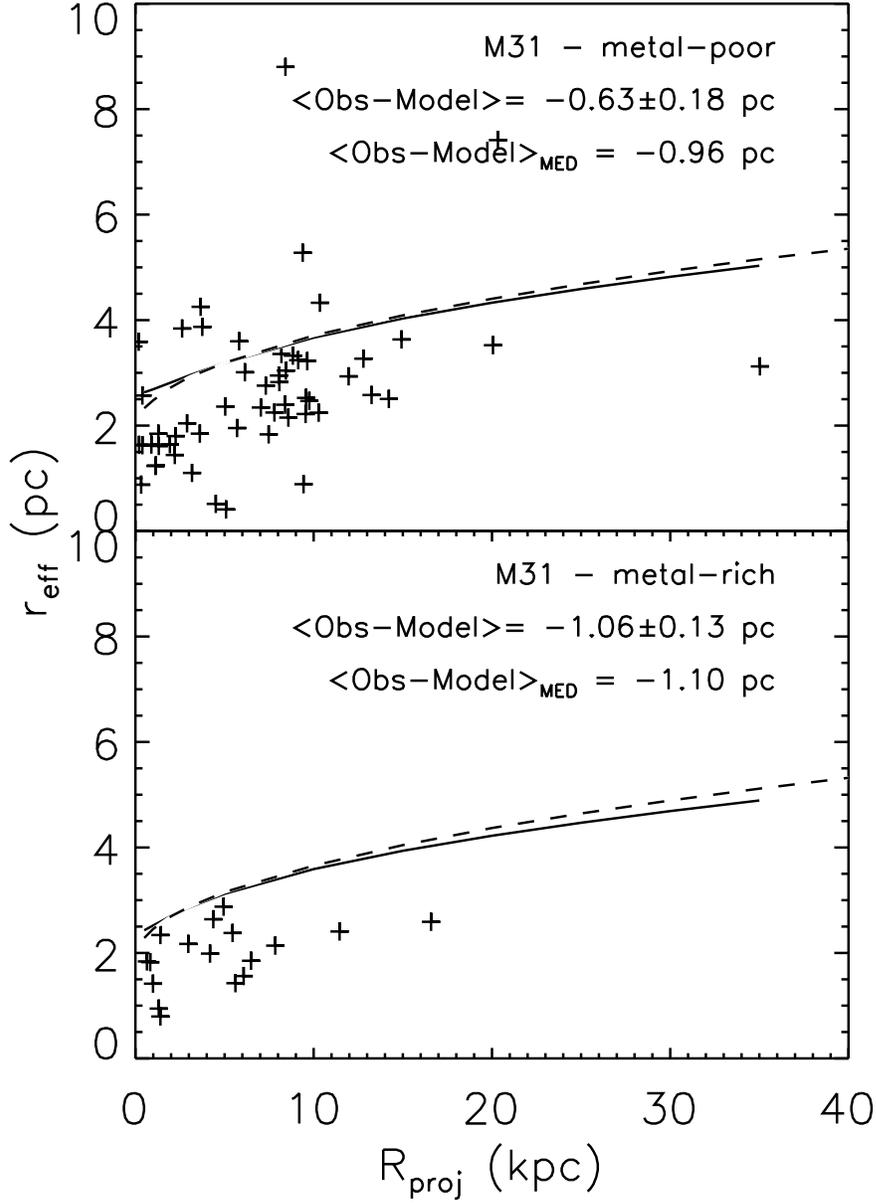}
\end{minipage}
\figcaption[]{\label{fig:rszm31}Size vs.\ projected galactocentric distance 
  for GCs in M31. In each panel, the solid and dashed lines show the mean 
  sizes obtained by projecting the Milky Way size-$R_{\rm proj}$ relation for
  King and de Vaucouleurs GC density profiles, respectively. 
}
\end{figure}

  In Fig.~\ref{fig:rszm31} we apply Eq.~(\ref{eq:szavg}) to the M31 GC system,
adopting the same size-$R$ relation as in the Milky Way. The top and bottom
panels show GC size vs.\ projected galactocentric distance for metal-poor
and metal-rich GCs, respectively.  As noted by other authors, there is a 
clear trend of GC size increasing with $R$ for the M31 GCs 
\citep{cram85,bhh02}. Note, however, that the samples of M31 GCs are likely 
to suffer from severe selection effects, making the spatial distributions 
uncertain \citep{barm00}.

  In each panel of Fig.~\ref{fig:rszm31}, the solid and dashed lines 
represent the size-$R$ relations obtained by projecting the Milky Way
relation (Eq.~\ref{eq:r_R_mw}) using King and de Vaucouleurs GC density
profiles with effective radii of 17$\arcmin$ and 30$\arcmin$. These 
effective radii are based on the data in \citet{barm00} and are
rather similar to those for the Milky Way GC system. The first
impression is that the projected Milky Way relation appears to 
overestimate the GC sizes in M31, but this may be partly due to the
fact that the Milky Way contains a number of GCs with large sizes which
shift the mean relation upwards. Such clusters are not detected in M31,
but could well have evaded detection especially near the center of the
galaxy where the bright background from the bulge and disk makes it
difficult to see low-surface brightness, extended objects. 

  As indicated in the Figure, the mean differences between observed and 
model sizes are $-0.63\pm0.18$ pc and $-1.06\pm0.13$ pc for the metal-poor 
and metal-rich GCs, respectively. Thus, in addition to the
systematic offset between modeled and observed relations, there is a 0.4 pc 
difference between metal-rich and metal-poor clusters which is formally 
unaccounted for by simple projection of the 
Milky Way size-$R$ relation. However, there are several outlying datapoints,
especially in the upper panel.
The median is less sensitive to such outliers, and if
we use the median instead of mean as an indicator of the difference
between model and observations (labeled 
$\langle{\rm Obs}-{\rm Model}\rangle_{\rm MED}$ 
in Fig.~\ref{fig:rszm31}) then the difference is reduced to 0.14 pc.
In summary, it seems quite likely that the size difference between the 
GC sub-populations in M31 may again be explained largely as a result 
of differences in the radial distributions, as in the Milky Way, 
although it would desirable to confirm this with size measurements for
a larger sample of clusters.

\subsection{Projected GC size vs.\ radius trends in the HST datasets}

\begin{figure}
\begin{minipage}{12cm}
\plotone{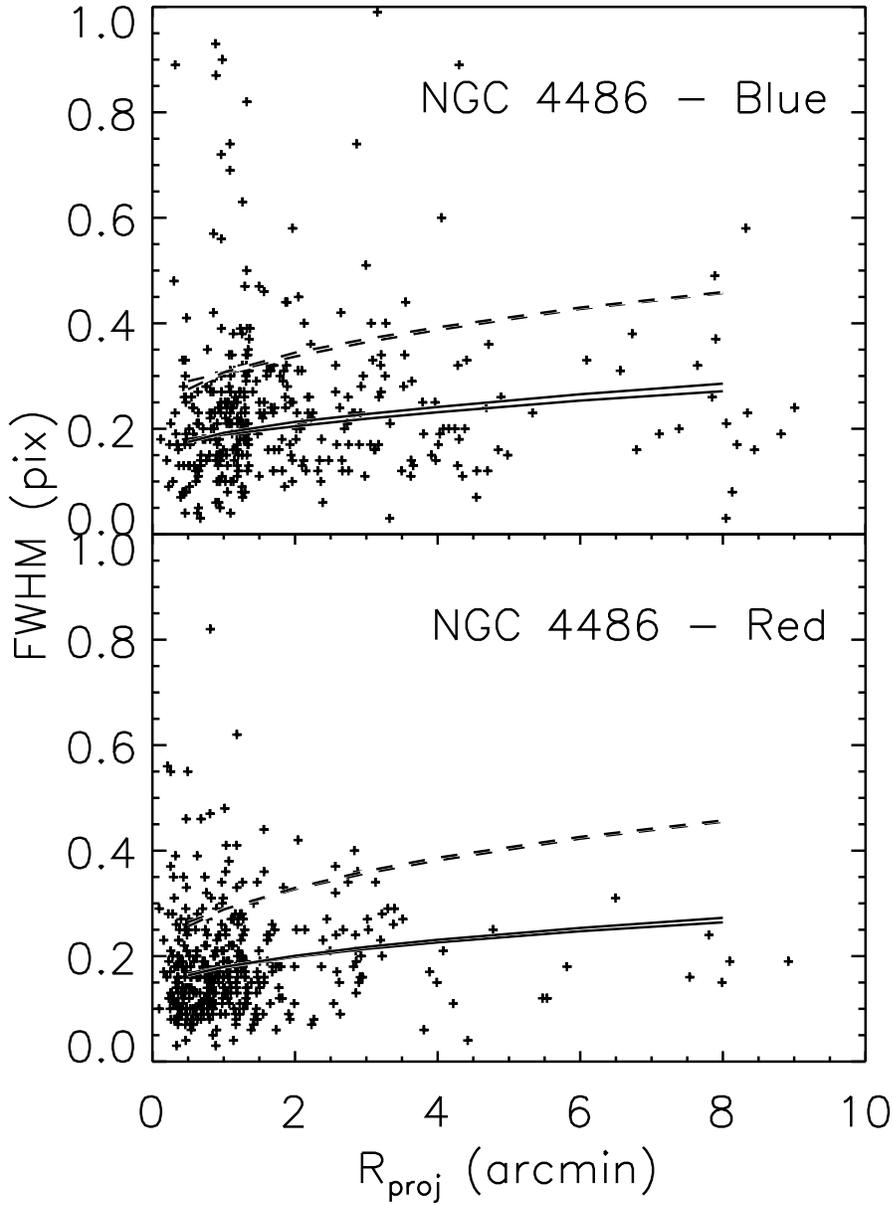}
\end{minipage}
\figcaption[]{\label{fig:rszfit}Size vs.\ projected galactocentric distance 
  for GCs in NGC~4486. The dashed curves show the mean sizes
  obtained by projecting the Milky Way size-$R_{\rm proj}$ relation for
  King and de Vaucouleurs GC density profiles. The solid curves represent
  a relation with half the slope of the Milky Way relation and a zero-point
  shifted downwards to roughly match the NGC~4486 data.
}
\end{figure}

  NGC~4472 and NGC~4486 are both located in the Virgo cluster, for which
we will assume a distance of 15 Mpc. NGC~4594 is somewhat closer, at a 
distance of about 9 Mpc \citep{ford96}.  At 15 Mpc, 1 WF pixel corresponds to 
7.3 pc.  With a 
conversion factor of 1.48 between the FWHM and half-light radii of the King 
models used to fit the cluster profiles, the Milky Way relation 
(Eq.~\ref{eq:r_R_mw}) then translates to 
${\rm FWHM} \, \approx \, (0.08 \, {\rm pixels}) \sqrt{R_{\rm 3D} / 0\farcm7} \, + \, 0.17$ pixels (where the scale radius of 3 kpc corresponds to
$0\farcm7$ at the assumed distance).  It is, however, not obvious 
just how to apply the Milky Way size-$R_{\rm 3D}$ relation to other galaxies.  
In Fig.~\ref{fig:rszfit} we display the size measurements for red and blue 
GCs in NGC~4486 as a function of $R_{\rm proj}$. The radial range, $10\arcmin$
or 44 kpc, is roughly similar to those for the Milky Way and M31 plots in
Fig.~\ref{fig:r_sz_mw} and Fig.~\ref{fig:rszm31}.
In each
panel, the mean size vs.\ $R_{\rm proj}$ relation obtained by projecting the 
Milky Way relation is shown with dashed lines, using Eq.~(\ref{eq:szavg}) and 
the King and de Vaucouleurs profiles from Fig.~\ref{fig:rp4486}. At
the scale and radial range of this figure, the two model GC surface
density profiles give very similar mean size vs. $R_{\rm proj}$ relations.

  While the sizes are measured on several WFPC2 pointings and systematic 
differences might be present from one pointing to another, it is
undeniable that the Milky Way relation applied directly 
provides a rather poor fit to the 
NGC~4486 data.  Part of the mismatch might be due to systematic errors in 
the PSF modeling, which could shift all 
datapoints up or down, but application of the Milky Way relation also 
gives a too steep slope. There is, of course, no particular 
reason why the size-$R$ relations should be the same in the Milky Way 
and in NGC~4486.  The solid lines in Fig.~\ref{fig:rszfit} show the 
size-$R_{\rm proj}$ 
relation obtained by projecting a size-$R_{\rm 3D}$ relation with half the 
slope of that in the Milky Way and shifted downwards by 0.05 pixel.  This 
provides a much better match to the observations, although the match
remains less than satisfactory in the outer regions.
Note that, instead of explicitly applying a 
shallower slope to the size-$R_{\rm 3D}$ relation, we could have normalized 
$R_{\rm 3D}$ to a 4 times larger scale radius. Although the physical basis
for the size-$R$ relation is poorly understood, it is interesting to note
that half-light radius of the red GC population in NGC~4486, $2\arcmin$ or 
9 kpc, is also roughly 3--4 times larger than that of the metal-rich
Milky Way GCs.  

\begin{figure}
\begin{minipage}{12cm}
\plotone{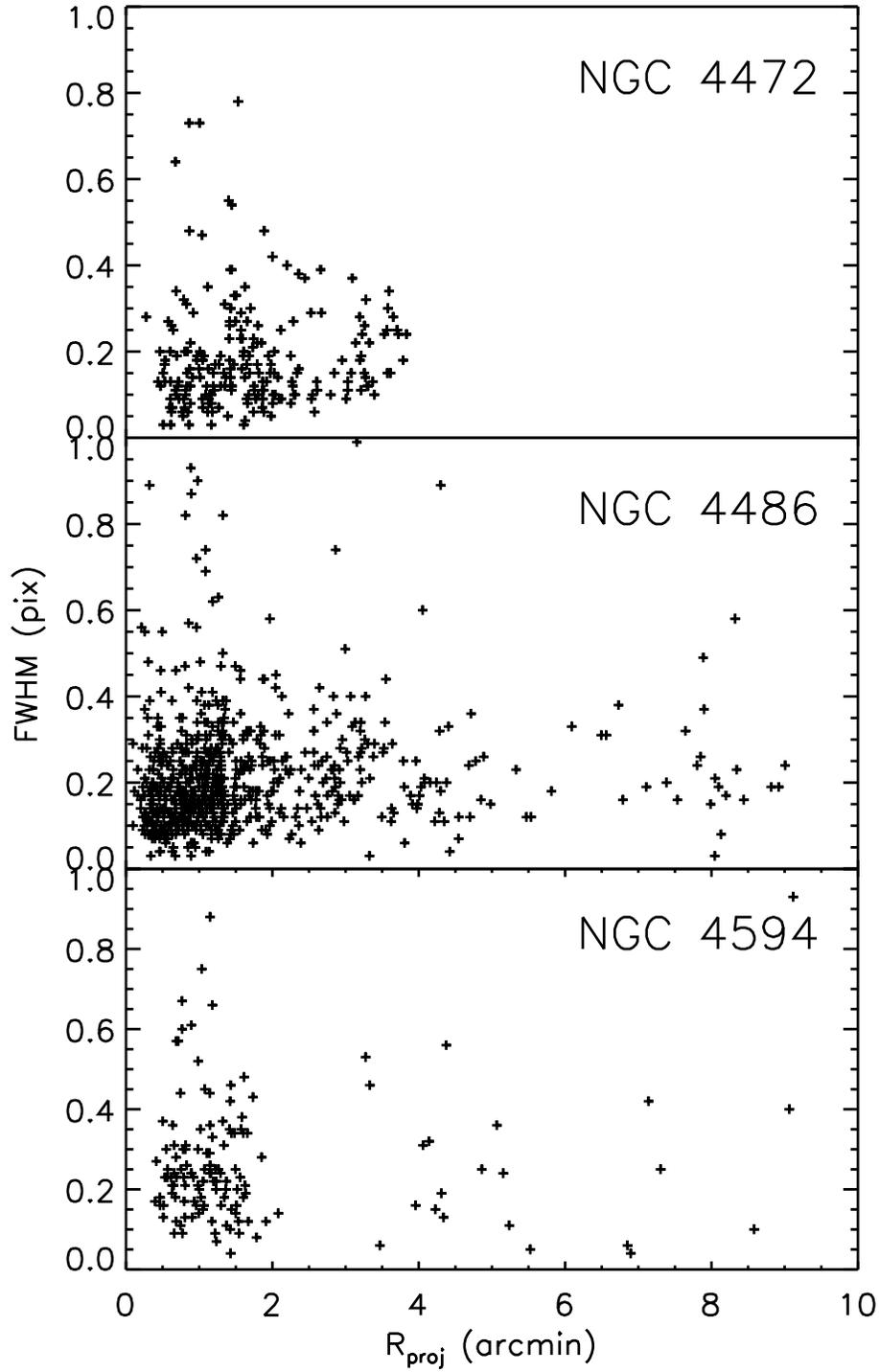}
\end{minipage}
\figcaption[]{\label{fig:rszall}Size vs.\ projected galactocentric distance 
  for GCs in NGC~4472, NGC~4486 and NGC~4594.
}
\end{figure}

\begin{figure}
\plotone{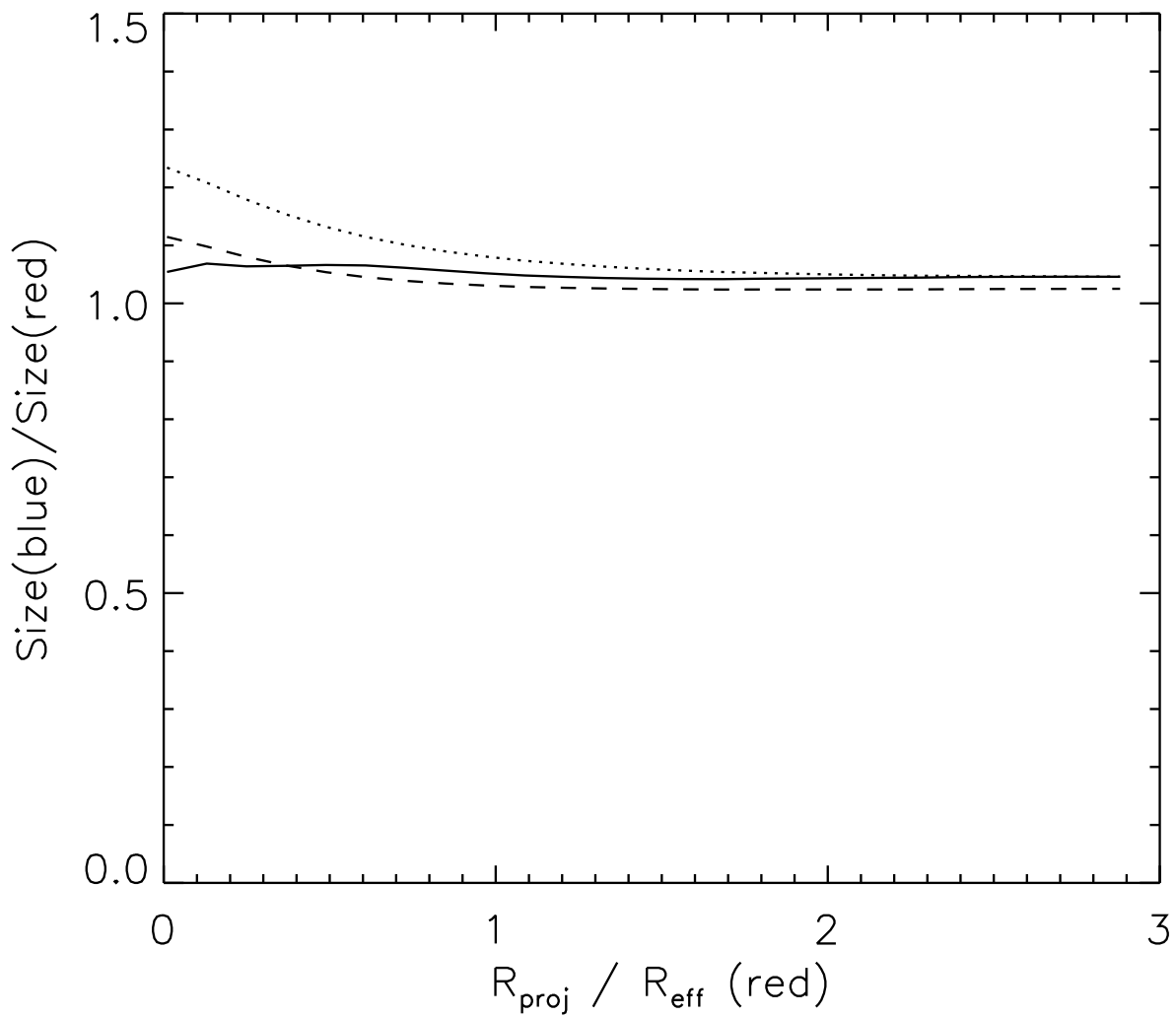}
\figcaption[]{\label{fig:ipar_s}Ratio of mean sizes of ``blue GCs'' and
  ``red GCs'' as a function of projected galactocentric distance. 
  Solid and dashed lines are for de Vaucouleurs and King profile fits to
  the surface density of the NGC~4486 GC system (Fig.~\ref{fig:rp4486}), 
  and the size-$R_{\rm 3D}$ relation used in Fig.~\ref{fig:rszfit}
  (Eq.~\ref{eq:sz_r_4486}).
  The dotted line is for King profiles whose effective radii differ by a
  factor of 4 (instead of $\sim2.3$).
}
\end{figure}

  The other HST datasets do not provide additional constraints on any 
size-$R$ relation for GCs, as illustrated in Fig.~\ref{fig:rszall}
where we compare GC size vs.\ $R_{\rm proj}$ for NGC~4472, NGC~4486 and 
NGC~4594. For the other two galaxies the radial range is
more limited and in NGC~4594 the number of clusters in the outer parts is 
very small (note that the NGC~4594 data cover a smaller radial range in kpc,
due to the smaller distance of this galaxy).  However, the behavior of GC 
size vs.\ $R_{\rm proj}$ in these 
galaxies does not appear to be significantly different from that in NGC~4486.
  Though the fit to the NGC~4486 data may be further improved by 
additional iterations on the input size-$R_{\rm 3D}$ relation, we will
adopt the half-slope, zero-point shifted relation in 
Fig.~\ref{fig:rszfit} for the following discussion, given as
\begin{equation}
  \mbox{FWHM} \, 
    = \, (0.12 \, {\rm pixels}) \, + \, (0.04 \, {\rm pixels}) \sqrt{R_{\rm 3D}/0\farcm7}
  \label{eq:sz_r_4486}
\end{equation}
  Fig.~\ref{fig:ipar_s} shows the ratio of the mean sizes of ``blue'' and 
``red'' GCs as a function of projected galactocentric distance, calculated 
according to Eq.~(\ref{eq:sz_r_4486}) and using the 
GC surface density profile fits in Fig.~\ref{fig:rp4486}.  The solid and 
dashed lines are for the de Vaucouleurs and King profiles, respectively.
In addition, the dotted line shows the ratio for King profiles for which
the effective radii of the blue and red GC surface density distributions
differ by a factor of 4 (instead of $\sim2.3$).  The projected radius 
$R_{\rm proj}$ is plotted in units of the effective radius of the red GC 
population.  Near the center, a difference of about 12\% in the mean sizes
is indeed produced for the King profiles, while the size difference
remains below 10\% for the de Vaucouleurs profiles. At larger radii the 
size difference decreases regardless of the choice of model.  While the 
details of Fig.~\ref{fig:ipar_s} clearly depend on many poorly constrained 
parameters, it provides another hint that projection effects might 
account, at least partially, for the differences between the projected size 
distributions of metal-rich and metal-poor GCs.

\subsection{Modeling the size distributions}

  A lot of useful information is lost by simply averaging over the size 
distributions and comparing mean sizes. Even if the average trends are 
reproduced, the detailed 
behavior of the observed distributions might not be properly accounted for.  
Instead, it is more illustrative to directly compare the predicted size 
\emph{distributions} with the observations.

  At a given projected distance $R_{\rm proj}$, the number of clusters 
$N(R_{\rm 3D})dR_{\rm 3D}$ in a small radial range $dR_{\rm 3D}$ around 
$R_{\rm 3D}$ is proportional 
to the right-hand side of Eq.~(\ref{eq:ns}).  It follows that the size
distribution, expressed as the number of clusters $N(r)dr$ in a small size 
bin $dr$ centered on $r$ can be expressed as
\begin{equation}
   N(r)dr \, = \, 
     \frac{\rho \, R_{\rm 3D}}{\sqrt{R_{\rm 3D}^2 - R_{\rm proj}^2}}
     \frac{dR_{\rm 3D}}{dr} \, dr
  \label{eq:nr}
\end{equation}
  where it is assumed that $r$ is a monotonic function of $R_{\rm 3D}$ so 
that the inverse function $R_{\rm 3D}(r)$ exists. Note also that the space
density of clusters, $\rho$, must be evaluated at the appropriate 
radius, $R_{\rm 3D}(r)$. 

\begin{figure}
\plottwo{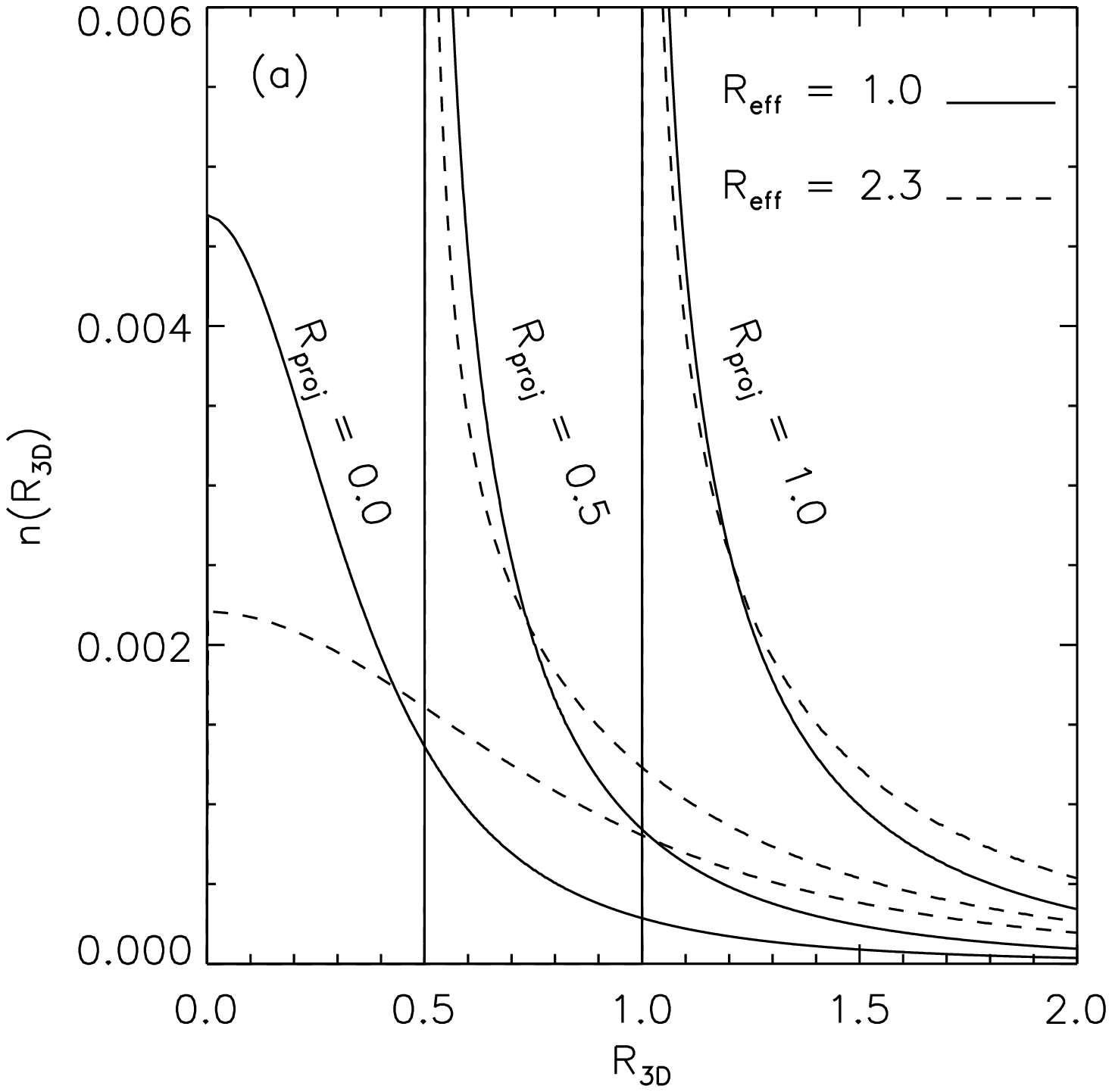}{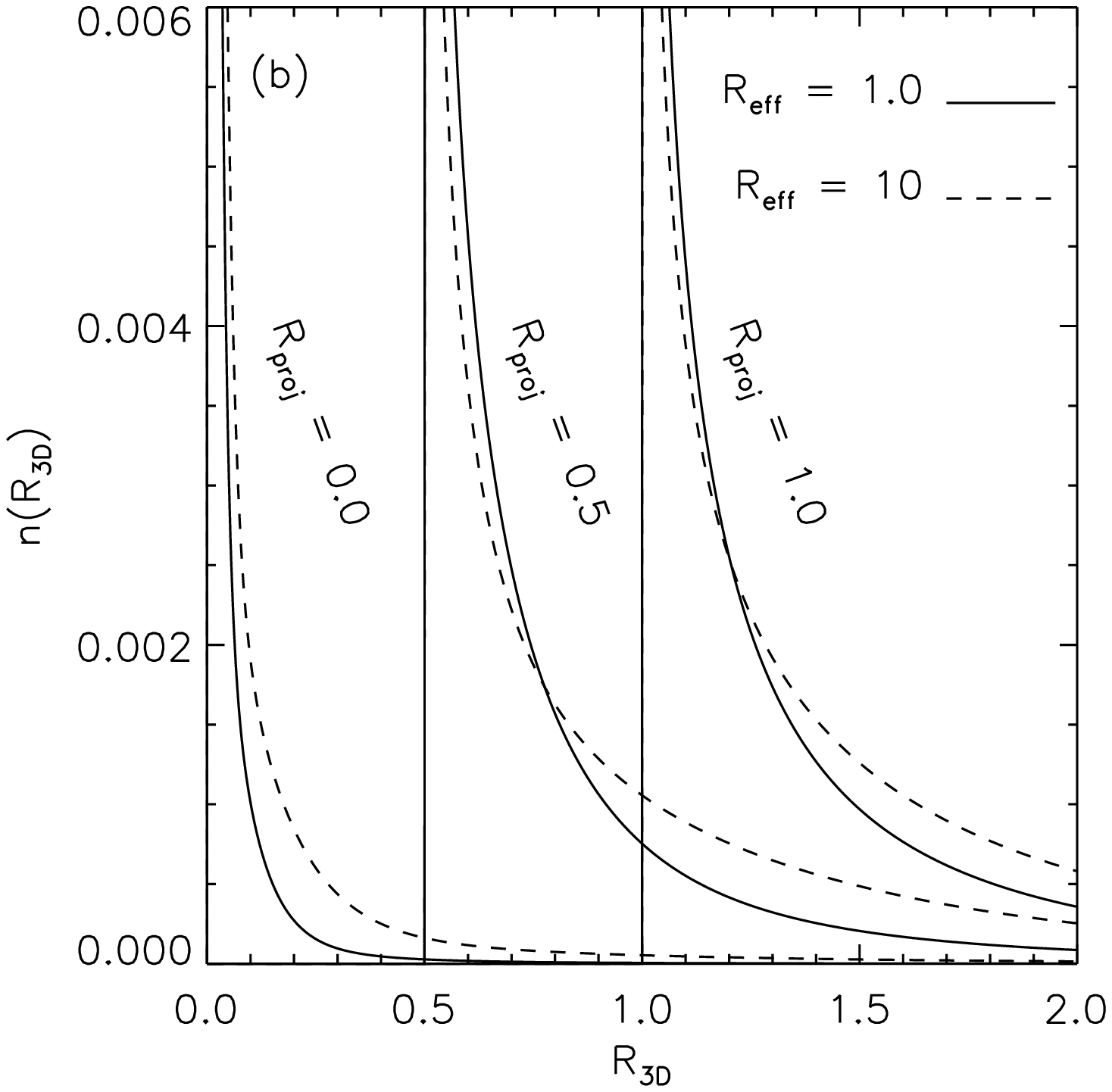}
\figcaption[]{\label{fig:nrdr}Distribution functions for 
  $n(R_{\rm 3D})$ at projected radii of 0.0, 0.5 and 1.0 for King models
  (a) and de Vaucouleurs laws (b). The solid lines represent models
  with effective radii of 1.0 while the dashed lines are for effective
  radii of 2.3 (King models) or 10 (de Vaucouleurs laws).
}
\end{figure}

  The distribution functions for $R_{\rm 3D}$ (Eq.~\ref{eq:ns}) at various 
projected radii are 
shown in Fig.~\ref{fig:nrdr} for King profiles (a) and de Vaucouleurs laws (b).
Each panel shows the $R_{\rm 3D}$ distribution functions for projected radii 
of $R_{\rm proj} = 0.0$, 0.5 and 1.0. The solid lines are for models with 
effective radii of 1.0 (both panels). Following the fits to the NGC~4486 GC 
system (Fig~\ref{fig:rp4486}), the dashed lines show King models with 
$\reff = 2.3$ (a) and de Vaucouleurs models with $\reff=10$ (b). For both 
models the $R_{\rm 3D}$ distributions are strongly peaked near $R_{\rm proj}$ 
and quite similar for models with different effective radii, except for the 
King models near the center. Hence, the most significant differences in
the size distributions of blue and red GCs are expected near the center,
and should be much more noticeable for King density profiles.

\begin{figure}
\plottwo{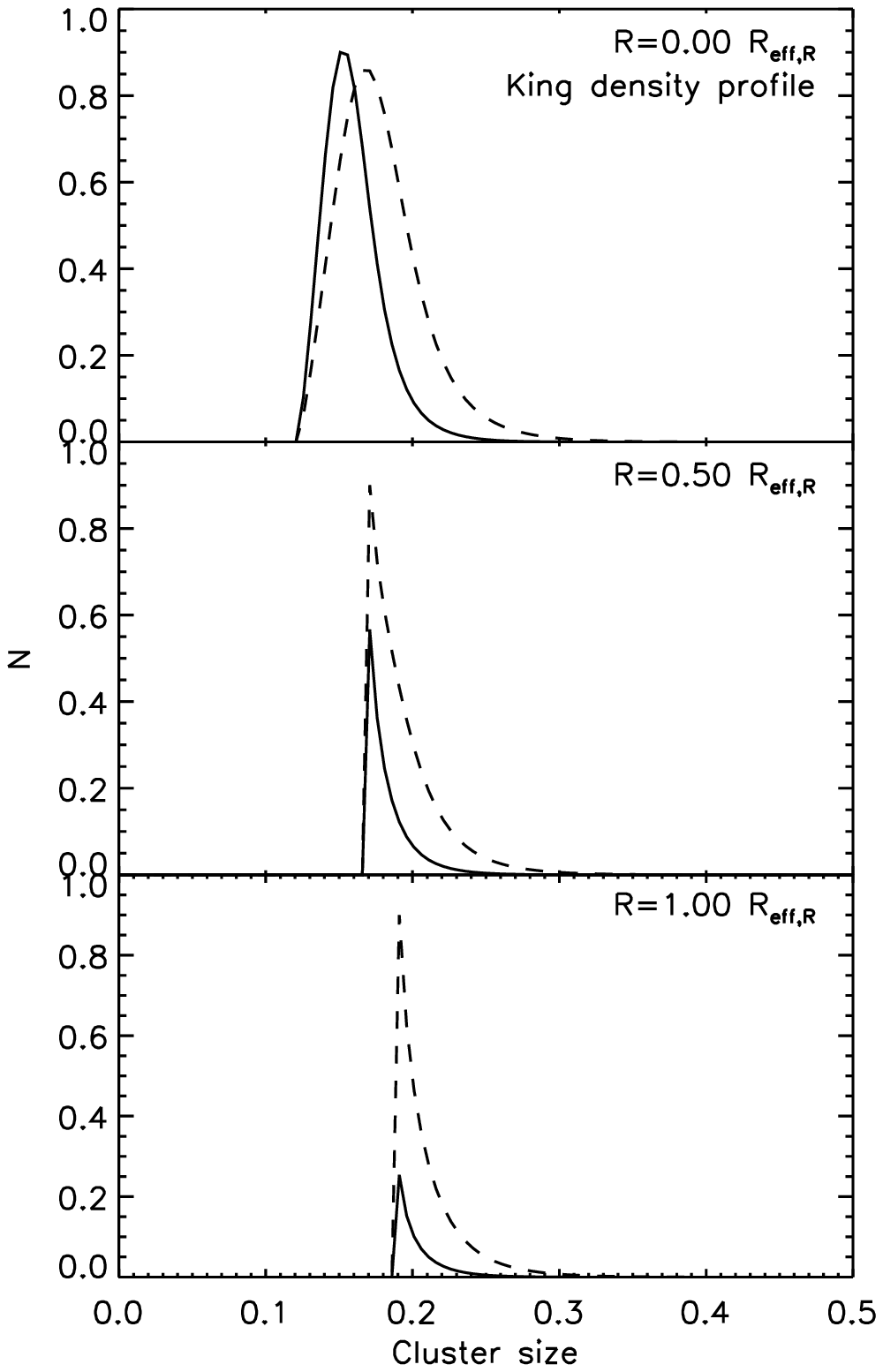}{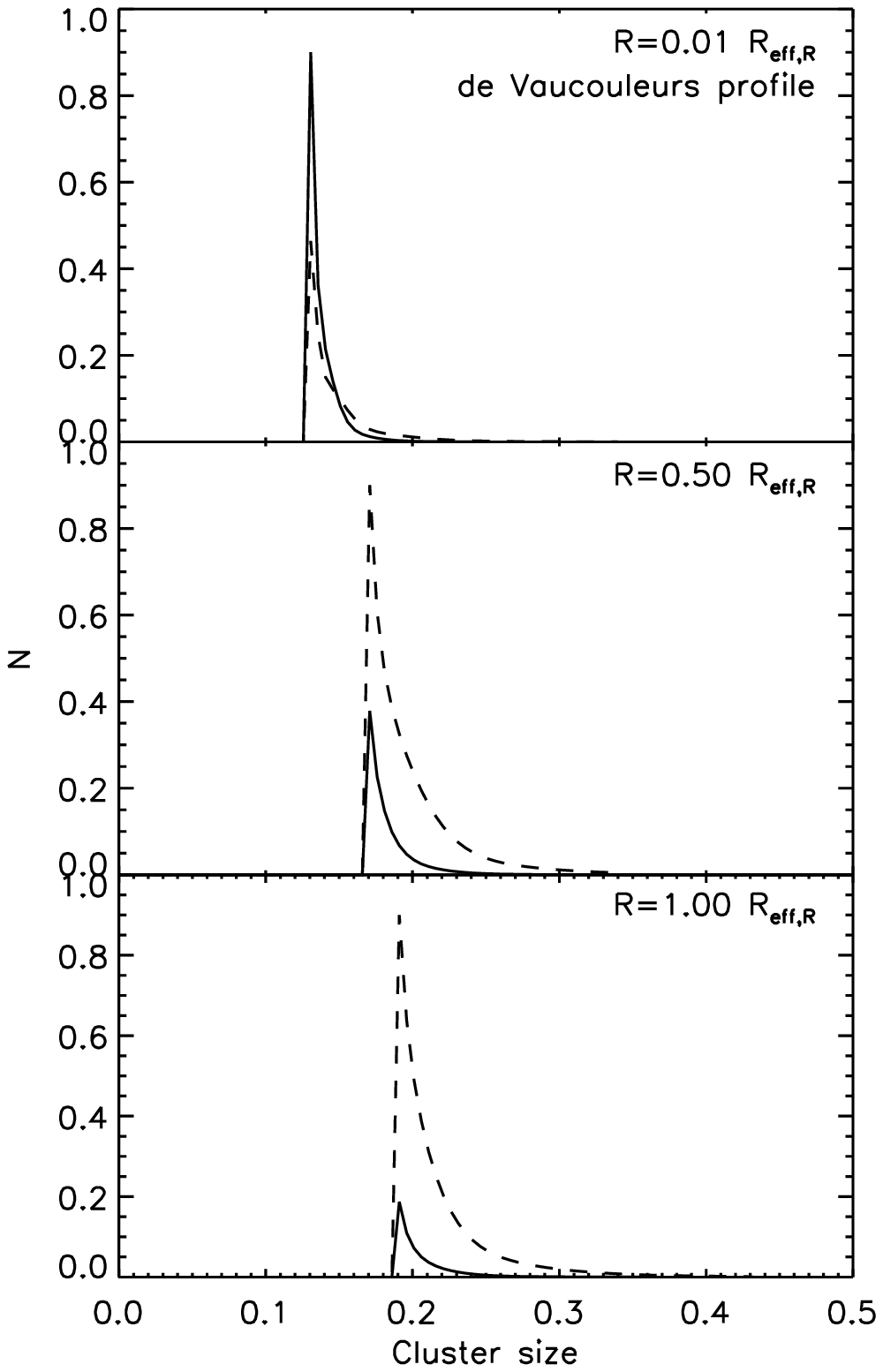}
\figcaption[]{\label{fig:szsim}Simulated size distributions for King (left)
  and de Vaucouleurs (right) density profiles and size-$R$ relation used in
  Fig.~\ref{fig:rszfit}. The solid and dashed curves represent the
  distributions obtained by using the fits to the radial density
  profiles of red and blue GCs, respectively, in NGC~4486
  (see Fig.~\ref{fig:rp4486}).
}
\end{figure}

  In Fig.~\ref{fig:szsim} we show the projected size distributions at the
same $R_{\rm proj}$ as in Fig.~\ref{fig:nrdr}, using the size-$R_{\rm 3D}$ 
relation for NGC~4486 from Fig.~\ref{fig:rszfit} (Eq.~\ref{eq:sz_r_4486}).  
As suspected, the King 
and de Vaucouleurs laws produce very different 
size distributions near the center. Even though the de Vaucouleurs law does 
produce a difference in the mean sizes (though smaller than for the King 
profiles), this is mainly due to a more extended ``tail'' in the size 
distribution of the blue clusters.  For the King profiles, on the other hand, 
the simulated size distributions show two clearly separated peaks
near the center, more akin to the observed distributions.  
Changing the constant term ($b$) in Eq.~(\ref{eq:r_R}) simply shifts both 
distributions horizontally, while changing the scale factor ($a$) changes 
the width of the two distributions and their separation. 

\begin{figure}
\plottwo{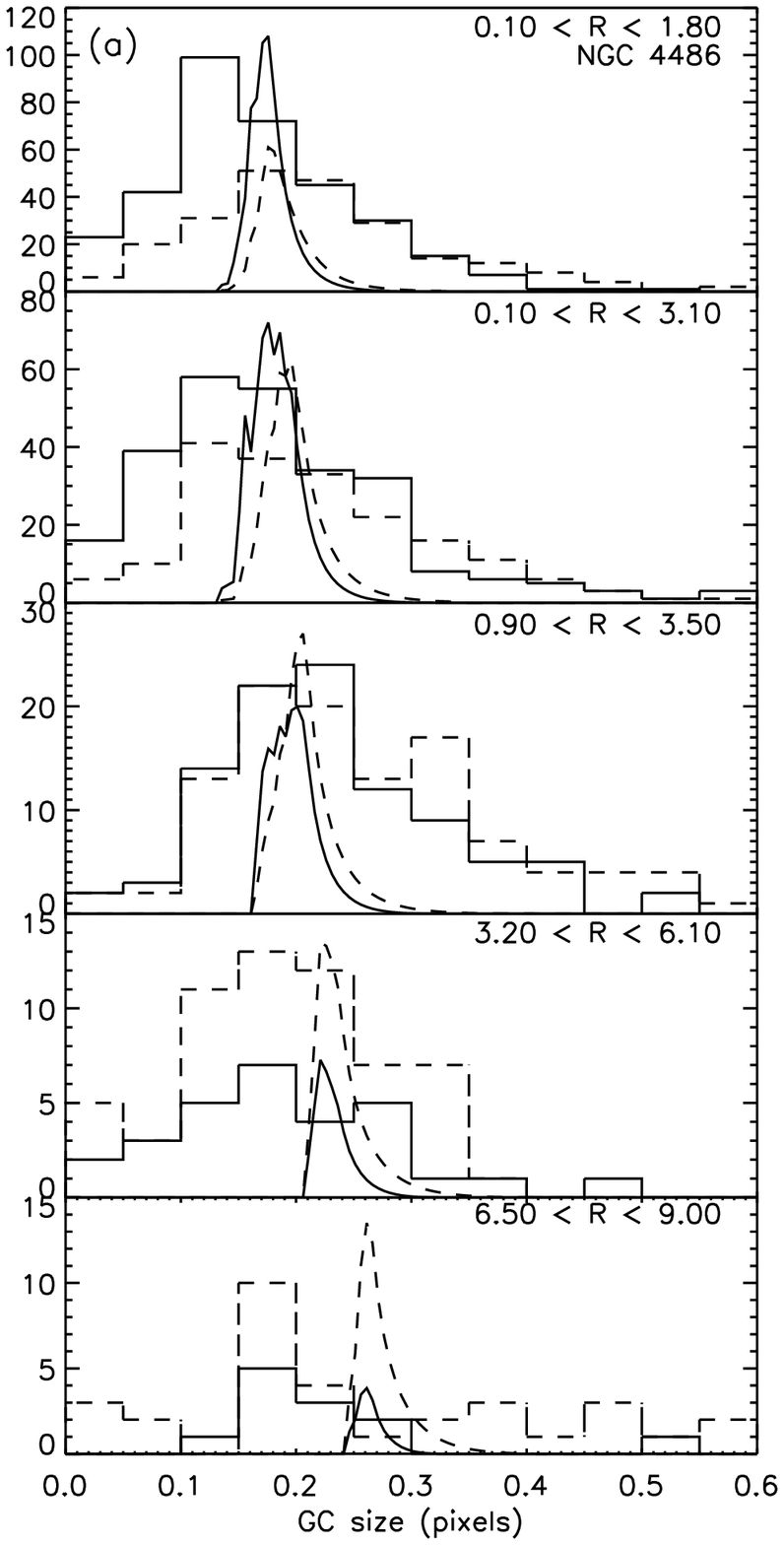}{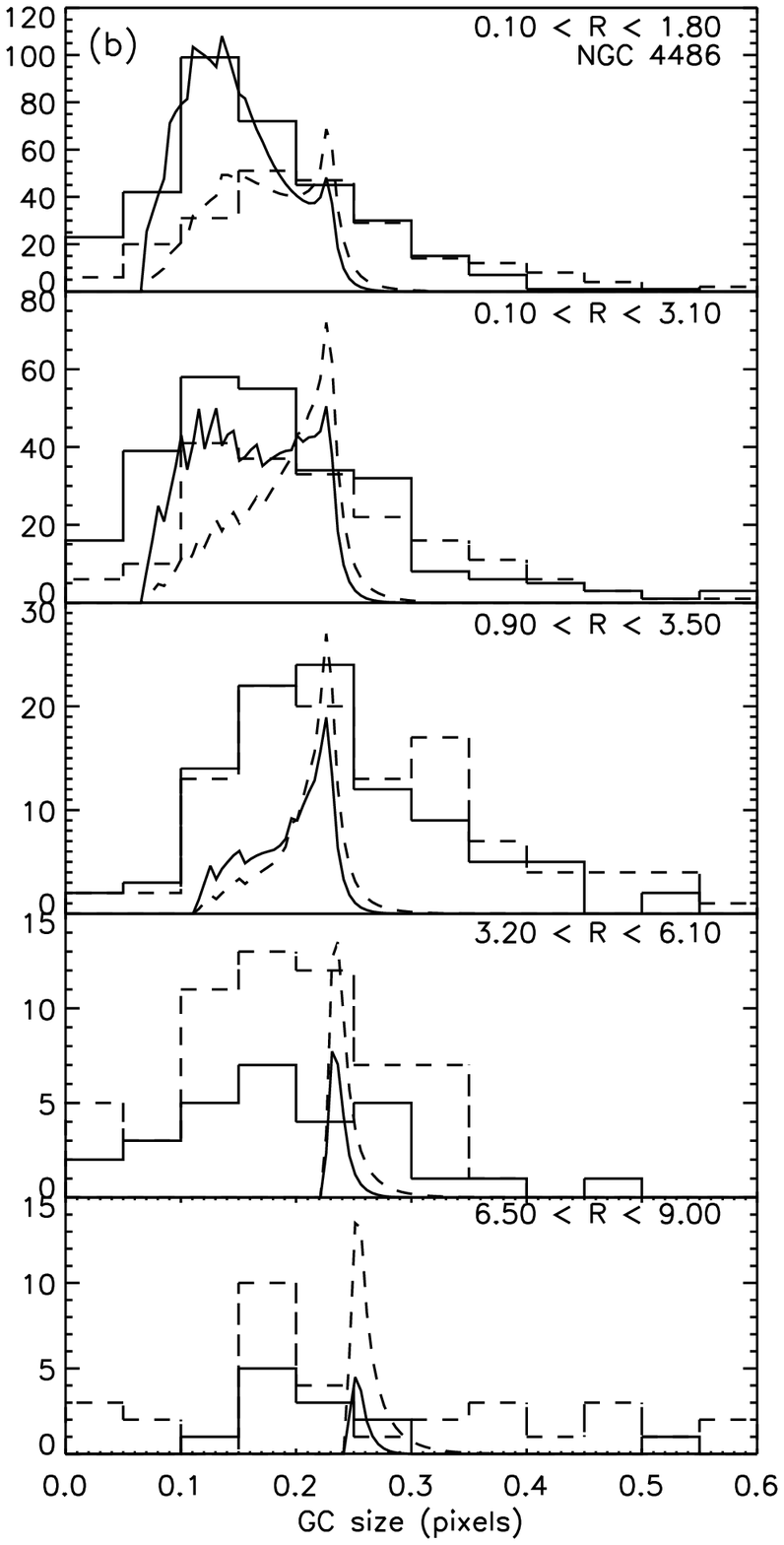}
\figcaption[]{\label{fig:szh4486k}Comparison of observed and simulated
  size distributions for red and blue GCs in NGC~4486 WFPC2
  pointings. For both observed and simulated distributions, the solid and 
  dashed curves represent ``red'' and ``blue'' GCs, respectively. 
  Panel (a) is for a square-root size-$R_{3D}$ relation while Panel (b)
  is for a relation defined by two linear segments. See text for details.
}
\end{figure}

  In Fig.~\ref{fig:szh4486k} we compare the observed size distributions 
in NGC~4486 with our simple model.  Panel (a) shows the size distributions
obtained by projecting the square-root size-$R_{\rm 3D}$ law 
(Eq.~\ref{eq:sz_r_4486}) at 15--30 discrete $R_{\rm proj}$ (hence the 
jagged appearance of some of the curves), spanning the radial range covered 
by each WFPC2 field.  The individual contributions were weighted by the 
azimuthal coverage of the WFPC2 field at the corresponding $R_{\rm proj}$. We 
have not attempted to normalize the model distributions to the observed 
numbers of clusters in each bin, although the relative numbers of red 
and blue clusters do change according to the assumed (King) surface density 
profiles. The observed and modeled number ratios N(red)/N(blue) are listed 
for each for each WFPC2 pointing in Table~\ref{tab:fittab} and agree
within the uncertainties, except perhaps for the outermost pointing
where deviations from the King profile may begin to become apparent. 

\begin{figure}
\plotone{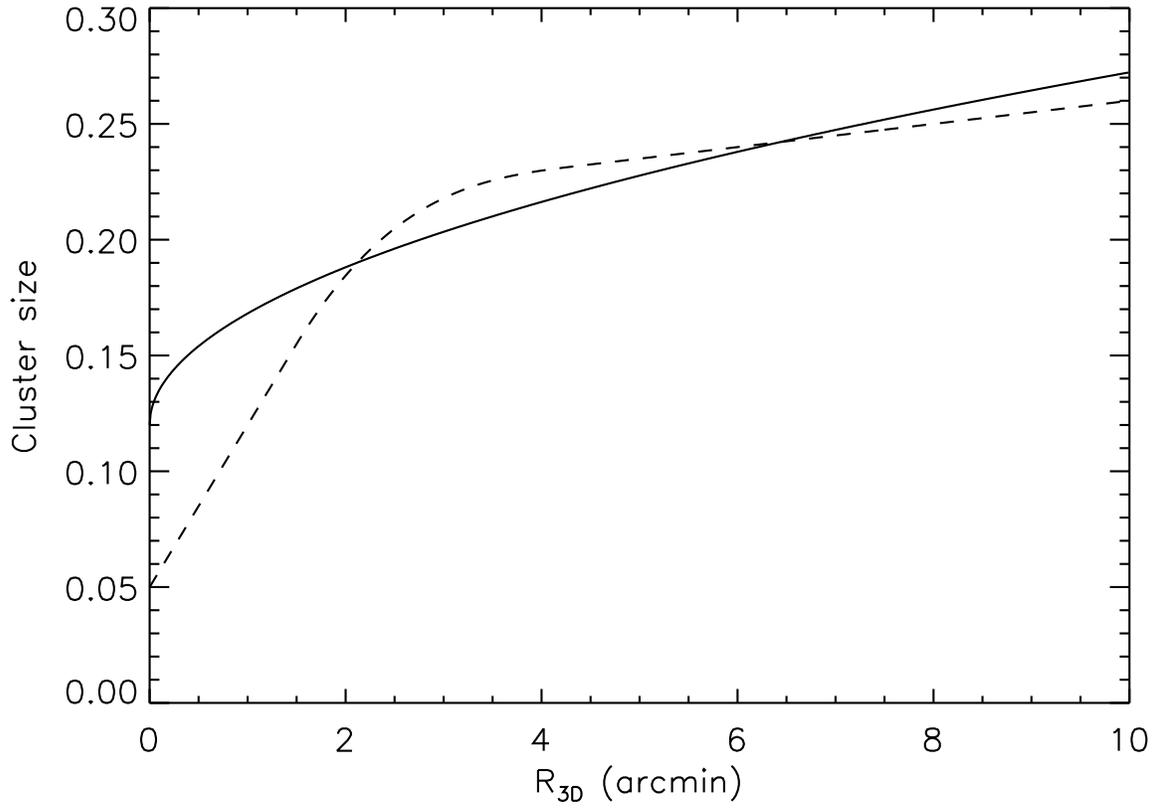}
\figcaption[]{\label{fig:sz_r_sim}The two different GC size vs. $R_{3D}$
  relations used in Fig.~\ref{fig:szh4486k}. The solid curve is a square 
  root relation while the dashed curve consists of two linear segments for 
  $R_{\rm 3D}<1\farcm5$ and $R_{\rm 3D}>4\farcm5$ and interpolation
  with a cubic spline in the range $1\farcm5<R_{\rm 3D}<4\farcm5$
  (Eq.~\ref{eq:r_sz2}).
}
\end{figure}

  While the simulated size distributions in Fig.~\ref{fig:szh4486k} (a) 
do show a significant difference between red and blue GCs, the offset between 
the two distributions is clearly not as pronounced as for the actual observed 
distributions. Thus, at
least for the simple square-root law, it is difficult to fully explain the
size difference as a projection effect. In order to obtain a wider
separation between the peaks of the two GC size distributions, we need 
a steeper relation between GC size and galactocentric distance 
in the central parts of the GC system. However, adopting a
steeper slope for the square-root law would be incompatible with
the relatively flat size-$R_{\rm 3D}$ trend observed at larger radii
(Fig.~\ref{fig:rszall}). We therefore need a curve which is steeper
at small radii compared to the square-root law, but flattens out at
large radii. One possibility is to use a relation consisting of two
linear segments:
\begin{equation}
  \mbox{FWHM} \, = \left\{ \begin{array}{lr} 
    a_1 R_{\rm 3D} + b_1 & \mbox{for } R_{\rm 3D} < R_1 \\
    a_2 R_{\rm 3D} + b_2 & \mbox{for } R_{\rm 3D} > R_2
    \end{array} \right.
   \label{eq:r_sz2}
\end{equation}
  with $R_2 > R_1$ and using a cubic spline to interpolate between the two 
linear segments in the interval $R_1 < R_{\rm 3D} < R_2$.  Just piecing
a relation together directly from two linear segments with an
abrupt change in slope at a certain radius will not work, because the
term $dR_{\rm 3D}/dr$ in Eq.~(\ref{eq:nr}) will then be discontinuous,
leading to a ``jump'' in the size distribution.
However, by choosing
an adequate separation between $R_1$ and $R_2$ the change in slope
will be gentle enough to avoid spurious effects in the resulting
size distribution.  After some
experimentation, we found that a reasonable match to the observed
size distributions was obtained for $a_1 = 0.07$ pixels arcmin$^{-1}$, 
$b_1 = 0.05$ pixels, $R_1 = 1\farcm5$, 
$a_2 = 0.005$ pixels arcmin$^{-1}$, $b_2 = 0.21$ pixels and
$R_2 = 4\farcm5$.  The curve corresponding to these coefficients
is compared with the square-root law in Fig.~\ref{fig:sz_r_sim} 
and in panel (b) of Fig.~\ref{fig:szh4486k} we 
show the corresponding size distributions.
By construction, the match to the 
observed size distributions is now much better in the inner pointing 
(uppermost panel), though the model now appears to actually overpredict 
the size difference in the range $0\farcm10<R<3\farcm10$.  The ``tail'' 
extending up to sizes $>0.5$ pixels in the observed
distributions is not reproduced in any of the model distributions, but
may well be due to intrinsic and/or measurement scatter in the GC
size distributions, which is not taken into account in our simple
model. Probably for the same reason, the simulated distributions in
the outer pointings are much more peaked than the observed ones.
Regardless, Fig.~\ref{fig:szh4486k} (b) shows that projection effects
can indeed produce size differences that are quite similar to the
observed ones. 

  Finally, we caution against over-interpreting Fig.~\ref{fig:szh4486k}.
It should be emphasized that the measured GC sizes are only a fraction 
of a WFPC pixel and any differences in mean size from one panel to another 
are probably only marginally significant.  For example, Table~\ref{tab:sztab} 
indicates a decrease in the mean cluster size from field NGC~4486-C to 
NGC~4486-D of about 0.06 pixels, which is almost certainly an artifact.  The 
random rms error on the size measurements is about 0.1 pixel \citep{lar01} 
and may account for a significant fraction of the widths of the observed 
size distributions, although the Milky Way data suggests that some intrinsic 
scatter is also present. 

\section{Summary and concluding remarks}

  We have re-examined the size distributions of globular cluster
sub-populations in three early-type galaxies with HST/WFPC2 imaging, as
well as those in the Milky Way and M31. As in previous studies, we find that 
the blue clusters are, on average, larger than the red ones. The difference
is, however, most pronounced in the central parts of the galaxies, and its 
statistical significance decreases strongly at larger radii. 

  In order to test whether the size difference might be a projection effect, 
resulting from a correlation between cluster size and galactocentric distance 
combined with different radial distributions of the GC subpopulations, we 
first fitted a square-root law to size vs.\ galactocentric distance 
($R_{\rm 3D}$) for metal-rich and metal-poor GCs in the Milky Way. Within the 
uncertainties, the two GC sub-populations in the Milky Way appear to follow 
the same size-$R_{\rm 3D}$ relation. For extragalactic GC systems, only 
projected size distributions can be observed. In general, the size 
distributions observed at a given projected galactocentric distance will 
depend on the relation between GC size and 3-D galactocentric distance, as 
well as the 3-D density profiles of the two GC populations.
These relations are poorly constrained by current data, but we have explored 
a limited number of analytical approximations that
seem to be adequate for the NGC~4486 GC system (for which the best data
are available). Specifically, we 
modeled the projected size distributions of GC subpopulations in
NGC~4486 by assuming de
Vaucouleurs $R^{1/4}$ and King profiles for the GC density profiles and 
projecting plausible three-dimensional size-$R_{\rm 3D}$ relations for the GC 
subpopulations. For the de Vaucouleurs profiles we fail to
reproduce the observed size differences, but for the King profiles we find 
that, with some fine-tuning of the input size--$R_{\rm 3D}$ relation, we are 
able to produce size distributions for the red and blue GC subpopulations that 
are quite similar to the observations. In our model, the size difference 
between red and blue GCs is expected to decrease strongly beyond 1 
effective radius of the GC system. This is consistent with the lack
of any size-color relation for GCs in NGC~5128 noted by \citet{har02}, 
as all of the clusters in their sample are located beyond 6 kpc (compared to 
an effective radius of the underlying stellar halo of 5.2 kpc
\citep{peng02}).

  Our approach clearly involves many simplifying assumptions, such as ignoring 
any intrinsic scatter in the GC sizes at any given galactocentric distance and
using weakly constrained input relations for GC size vs.\ $R_{\rm 3D}$.
It nevertheless seems likely that projection effects might indeed account for
a significant fraction, if not all, of the observed differences between the
size distributions of GC sub-populations. Thus, at this point it may be
unnecessary to resort to explanations involving different formation
and/or destruction
mechanisms. Since size differences have now been observed in many galaxies
and the radial distributions of blue and red GCs generally appear to be
different, the globular systems in those galaxies probably follow similar 
size-$R$ relations. It cannot be ruled out that intrinsic
size differences exist in some cases. Indeed, we know of some cases of
clusters with unusually large sizes, such as the faint Palomar-type clusters
in the Milky Way halo and the ``faint fuzzy'' clusters in the disks of the 
nearby lenticular galaxies NGC~1023 and NGC~3384 \citep{lb00,bl02}. These
objects have effective radii that are a factor of 3--4 larger than those of 
``normal'' star clusters, and might have formed under different conditions.  

\acknowledgments

This work has benefited from discussions and correspondence with many of 
our colleagues, including D.\ Forbes, J.\ Strader and J.\ Cohen. The
presentation has also benefitted from the useful comments of an anonymous
referee.  We acknowledge support by National Science Foundation grants 
AST9900732 and AST0206139.

\onecolumn

\begin{deluxetable}{lccccc}
\tablecaption{ \label{tab:sztab}Mean sizes of blue and red GCs (in WF pixels)
 in the various WFPC2 pointings. $P$ is the probability from a
 Kolmogorov-Smirnoff test that the size distributions of blue and red
 GCs are drawn from the same parent distribution.
}
\tablehead{Field  & Range & \multicolumn{3}{c}{Mean FWHM (WF pixels)} & $P$ \\
                  &       & Blue & Red & All
}
\startdata
NGC 4472-A & $0\farcm3<R<2\farcm0$ & $0.186\pm0.014$ & $0.158\pm0.014$ & $0.169\pm0.010$ & 0.025 \\ 
NGC 4472-B & $1\farcm4<R<3\farcm9$ & $0.205\pm0.021$ & $0.204\pm0.021$ & $0.204\pm0.015$ & 0.334 \\ 
NGC 4472-C & $1\farcm3<R<3\farcm9$ & $0.214\pm0.023$ & $0.156\pm0.013$ & $0.183\pm0.013$ & 0.114 \\ 
NGC 4486-A & $0\farcm1<R<1\farcm8$ & $0.236\pm0.010$ & $0.170\pm0.006$ & $0.197\pm0.006$ & 0.000 \\ 
NGC 4486-B & $0\farcm1<R<3\farcm1$ & $0.230\pm0.011$ & $0.176\pm0.006$ & $0.199\pm0.006$ & 0.001 \\ 
NGC 4486-C & $0\farcm8<R<3\farcm5$ & $0.287\pm0.016$ & $0.230\pm0.011$ & $0.261\pm0.010$ & 0.029 \\ 
NGC 4486-D & $3\farcm2<R<6\farcm1$ & $0.212\pm0.020$ & $0.186\pm0.019$ & $0.203\pm0.015$ & 0.999 \\ 
NGC 4486-E & $6\farcm5<R<9\farcm0$ & $0.276\pm0.032$ & $0.317\pm0.065$ & $0.289\pm0.029$ & 0.968 \\ 
NGC 4594-A & $0\farcm4<R<2\farcm1$ & $0.276\pm0.018$ & $0.236\pm0.020$ & $0.257\pm0.014$ & 0.029 \\ 
NGC 4594-B & $3\farcm3<R<5\farcm6$ & $0.162\pm0.042$ & $0.271\pm0.068$ & $0.212\pm0.039$ & 0.654 \\ 
NGC 4594-C & $6\farcm9<R<9\farcm3$ & $0.267\pm0.140$ & $0.350\pm0.247$ & $0.297\pm0.113$ & 0.377 \\ 
\enddata
\end{deluxetable}

\begin{deluxetable}{lcc}
\tablecaption{ \label{tab:fittab}Relative numbers of clusters in
  NGC~4486 pointings, shown for both observations and simulations (see
  text for details).}
\tablehead{Field  & \multicolumn{2}{c}{N(red)/N(blue)} \\
                  &    Observed & Model 
}
\startdata
NGC~4486-A & $1.50\pm0.13$ & 1.65 \\
NGC~4486-B & $1.39\pm0.13$ & 1.34 \\
NGC~4486-C & $0.90\pm0.13$ & 0.92 \\
NGC~4486-D & $0.49\pm0.11$ & 0.46 \\
NGC~4486-E & $0.38\pm0.13$ & 0.23 \\
\enddata
\end{deluxetable}
\clearpage

\end{document}